\documentclass[a4paper]{article}
\usepackage{amsmath}
\usepackage{amsthm}
\usepackage{amssymb}
\usepackage{amsfonts}
\usepackage[latin1]{inputenc}
\usepackage{alltt}
\usepackage{xspace}
\usepackage{graphics}
\usepackage{listings}
\usepackage{multicol}
\usepackage{a4wide}

\newcommand{\keyword}[1]{\textbf{\sffamily #1}}

\def\mob{\textsc{Mob}\xspace}

\def\classk{\keyword{class}\xspace}
\def\agentk{\keyword{agent}\xspace}
\def\servicek{\keyword{service}\xspace}
\def\implementsk{\keyword{provides}\xspace}
\def\requiresk{\keyword{requires}\xspace}

\def\hostk{\keyword{host}\xspace}

\def\ifk{\keyword{if}\xspace}
\def\elsek{\keyword{else}\xspace}
\def\whilek{\keyword{while}\xspace}
\def\breakk{\keyword{break}\xspace}
\def\execk{\keyword{exec}\xspace}
\def\exitk{\keyword{exit}\xspace}
\def\bindk{\keyword{bind}\xspace}
\def\gok{\keyword{go}\xspace}

\def\forkk{\keyword{fork}\xspace}
\def\waitk{\keyword{wait}\xspace}
\def\notifyk{\keyword{notify}\xspace}
\def\joink{\keyword{join}\xspace}
\def\returnk{\keyword{return}\xspace}
\def\newk{\keyword{new}\xspace}
\def\maink{\keyword{main}\xspace}
\def\lockk{\keyword{lock}\xspace}
\def\unlockk{\keyword{unlock}\xspace}

\def\truek{\keyword{true}}
\def\falsek{\keyword{false}}

\def\selfk{\keyword{self}\xspace}

\def\nullk{\keyword{null}\xspace}
\def\nullr{\mathmob{null}\xspace}

\def\zeroka{\keyword{0}_\calA\xspace}
\def\zerokt{\keyword{0}_T\xspace}

\def\Var{\text{Var}\xspace}
\def\Class{\text{Class}\xspace} 
\def\Instruction{\text{Instruction}\xspace} 
\def\InstructionSeq{\text{InstructionSeq}\xspace} 
\def\Method{\text{Method}\xspace} 
\def\Def{\text{Definition}\xspace}
\def\Code{\text{Code}\xspace}
\def\ServiceId{\text{Service}\xspace} 
\def\MethodId{\text{MethodId}\xspace}
\def\Expression{\text{Expression}\xspace} 
\def\Heap{\text{Heap}\xspace}
\def\HeapRef{\text{HeapRef}\xspace}
\def\ThreadRef{\text{ThreadRef}}
\def\Agent{\text{Agent}\xspace} 
\def\AgentName{\text{AgentKey}\xspace} 
\def\NS{\text{NameResolver}\xspace} 
\def\ANS{\text{AgentNameResolver}\xspace} 
\def\SNS{\text{ServiceNameResolver}\xspace} 
\def\RunningThread{\text{RunningThread}\xspace}
\def\SuspendedThread{\text{SuspendedThread}\xspace}
\def\Closure{\text{Closure}\xspace}
 
\def\Type{\text{Type}\xspace} 
\def\Bool{\text{Bool}\xspace}
\def\Int{\text{Int}\xspace}

\def\String{\text{String}\xspace}
\def\Host{\text{Host}\xspace}
\def\Environment{\text{Bindings}\xspace}
\def\Network{\text{Network}\xspace}
\def\Constant{\text{Constant}\xspace}
\def\Value{\text{Value}\xspace}

\def\Stack{\text{CodeStack}\xspace}
\def\Pool{\text{Pool}\xspace}

\def\evalseq{\textsl{evalSeq}\xspace}
\def\evalone{\textsl{eval}\xspace}
\def\copyseq{\textsl{copySeq}\xspace}
\def\copyone{\textsl{copy}\xspace}
\def\access{\textsl{hasAccess}\xspace}
\def\locked{\textsl{tryLock}\xspace}
\def\unlocked{\textsl{tryUnlock}\xspace}
\def\codeOf{\textsl{code}\xspace}
\def\codein{\textsl{codeIn}\xspace}
\def\codeclo{\textsl{codeCollect}\xspace}
\def\notify{\textsl{notify}\xspace}
\def\run{\textsl{run}\xspace}
\def\launchT{\textsl{launch}\xspace}

\def\fresh{\text{fresh}\xspace}

\def\calA{\mathmob{A}}
\def\calR{\mathmob{R}}
\def\ns{\mathmob{R}}
\def\calN{\mathmob{N}}

\def\grmeq{\quad::=\quad}
\def\grmor{~~\mid~~}
\def\conc{\mid}

\newcommand{\rulename}[1]{[\textsc{#1}]\xspace}

\newcommand{\reduces}{\rightarrow}              

\def\cnrule#1#2#3{
  \bigskip
  \begin{small}
    \noindent
    #1\quad
    $\mathmob{#2} \reduces \mathmob{#3}$ 
  \end{small}
  \bigskip
}

\def\ucnrule#1#2#3{
  \bigskip
  \begin{small}
    \noindent
    #1\\
    $\mathmob{#2} \reduces \mathmob{#3}$ 
  \end{small}
  \bigskip
}

\def\incnrule#1#2{
     \mathmob{#1} \reduces \mathmob{#2}
}

\def\prule#1#2#3#4{
  \bigskip
  \begin{small}
  \noindent
  #1\quad
  \begin{tabular}{c}
    $\mathmob{#2}$ \\
    \hline
    $\mathmob{#3} \reduces \mathmob{#4}$ 
  \end{tabular}
  \end{small}
  \bigskip
}

\def\uprule#1#2#3#4{
  \bigskip
  \begin{small}
  \noindent
  #1\\
  \begin{tabular}{c}
    $\mathmob{#2}$ \\
    \hline
    $\mathmob{#3} \reduces \mathmob{#4}$ 
  \end{tabular}
  \end{small}
  \bigskip
}

\def\pnrule#1#2#3#4{
  \bigskip
  \begin{small}
  \noindent
  #1\quad
  \begin{tabular}{c}
    $\mathmob{#2}$ \\
    \hline
    $\mathmob{#3} \reduces$ \\ 
    \quad $\mathmob{#4}$ 
  \end{tabular}
  \end{small}
  \bigskip
}

\def\ppnrule#1#2#3#4#5{
  \bigskip
  \begin{small}
  \noindent
  #1\quad
  \begin{tabular}{c}
    $\mathmob{#2}$ \\
    $\mathmob{#3}$ \\
    \hline
    $\mathmob{#4} \reduces$ \\ 
    \quad $\mathmob{#5}$ 
  \end{tabular}
  \end{small}
  \bigskip
}

\def\upnrule#1#2#3#4#5{
  \bigskip
  \begin{small}
  \noindent
  #1\\
  \begin{tabular}{c}
    $\mathmob{#2}$ \\
    $\mathmob{#3}$ \\
    \hline
    $\mathmob{#4} \reduces \mathmob{#5}$ 
  \end{tabular}
  \end{small}
  \bigskip
}

\def\uppnrule#1#2#3#4#5{
  \bigskip
  \begin{small}
  \noindent
  #1\\
  \begin{tabular}{c}
    $\mathmob{#2}$ \\
    $\mathmob{#3}$ \\
    \hline
    $\mathmob{#4} \reduces$ \\ 
    \quad $\mathmob{#5}$ 
  \end{tabular}
  \end{small}
  \bigskip
}

\def\ppcnnrule#1#2#3#4#5#6{
  \bigskip
  \begin{small}
  \noindent
  #1\\
  \begin{tabular}{c}
    $\mathmob{#2}$ \\
    $\mathmob{#3}$ \\ 
   \hline
    $\mathmob{#4} \reduces$ \\ 
    \quad $\mathmob{#5}$\\
    \quad $\mathmob{#6}$ 
  \end{tabular}
\end{small}
  \bigskip
}

\def\pppcnnrule#1#2#3#4#5#6#7#8{
  \bigskip
  \begin{small}
  \noindent
  #1\\
  \begin{tabular}{ll}
    $\mathmob{#2}$ \\
    $\mathmob{#3}$ \\ 
    $\mathmob{#4}$ \\ 
    \hline
    $\mathmob{#5} \reduces$ \\ 
    \quad $\mathmob{#6}$ \\
    \quad $\mathmob{#7}$ \\ 
    \quad $\mathmob{#8}$ 
  \end{tabular}
\end{small}
\bigskip
}

\def\crule#1#2#3{
  \begin{small}
    \noindent
    {#1} \quad
    $\mathmob{#2}\ \equiv\ \mathmob{#3}$
  \end{small}
}

\def\pcrule#1#2#3#4{
\bigskip
  \begin{small}
    \noindent
    {#1} \quad
    \begin{tabular}{c}
      $\mathmob{#2}$ \\ 
      \hline 
      $\mathmob{#3}\ \equiv\ \mathmob{#4}$
    \end{tabular} 
  \end{small}
\bigskip
}

\def\pcrrule#1#2#3#4{
\bigskip
  \begin{small}
    \noindent
    {#1} \quad
    \begin{tabular}{c}
      $\mathmob{#2}$ \\ 
      \hline 
      $\mathmob{#3}\ \reduces\ \mathmob{#4}$
    \end{tabular} 
  \end{small}
\bigskip
}

\def\macrorule#1#2#3{
  \begin{small}
  \noindent
  {#1}\\ 
    $\mathmob{#2}\ \eqdef \mathmob{#3}$
  \end{small}
}

\lstdefinelanguage{mob}
{
morekeywords={requires, class, agent, service, provides, mobs, go, main,
    return, join, lock, unlock, if, else, while, for, in, break,
    switch, case, exit, new, fork, bind, host, exec, self, null, public},
sensitive=false,
morecomment=[l]{//},
morestring=[b]"
}

\def\intt{\textsl{int}\xspace}
\def\stringt{\textsl{string}\xspace}
\def\booleant{\textsl{bool}\xspace}

\def\threadt{\textsl{thread}\xspace}

\def\typeoff{\textsl{typeOf}\xspace}
\def\typeof#1{\typeoff(\mathmob{#1})}

\newenvironment{greduce}{
	\begin{small}
	\begin{eqnarray*}
}{
	\end{eqnarray*}
	\end{small}
        \vspace{-11pt}
}

\lstdefinelanguage{tyco}
{morekeywords={export, import, from, new, def, in, let, if, then, else, case, of},
sensitive=false,
morecomment=[l]{--},
morestring=[b]''
}

\def\conc{\mid}
\newcommand{\dom}{\mathit{dom}}

\newcommand{\ptyco}{\mbox{T\hspace{-.4ex}\raisebox{-.2ex}{y}CO}\xspace}
\newcommand{\dityco}{\ptyco}

\newcommand{\lsdpi}{\mbox{LSD}\pi}

\def\at#1#2{{#1}\sarroba{#2}}

\newcommand{\sarroba}{@}

\def\mob{\textsc{Mob}\xspace}
\def\mobam{\textsc{MobAM}\xspace}

\def\NS{\text{NameService}\xspace}
\def\ANS{\text{AgentNameService}\xspace}
\def\SNS{\text{ServiceNameService}\xspace}

 \def\access{\textsf{tryAccess}\xspace}
 \def\locked{\textsf{tryLock}\xspace}
 \def\unlocked{\textsf{tryUnlock}\xspace}

\def\fresh{\text{fresh}\xspace}

\def\calA{{\mathmob{A}}}
\def\calR{{\mathmob{R}}}
\def\ns{{\mathmob{R}}}
\def\Ans{\mathmob{ANS}\xspace}
\def\Sns{\mathmob{SNS}\xspace}
\def\calN{{\mathmob{N}}}

\def\grmeq{\quad::=\quad}
\def\grmor{~~\mid~~}
\def\conc{\mid}

\def\mathmob#1{\mathtt{#1}}

\def\dom{\textsl{dom}\xspace}

\def\accat#1{\acc@{#1}}

\def\eqdef{\;\stackrel{\text{def}}{=}\;}

\newenvironment{step}{\begin{small}\begin{eqnarray*}}{\end{eqnarray*}\end{small}}

\title{The \mob core language and abstract machine (rev 0.2)}

\author{Herv\'e Paulino$^1$ and  Lu\'is Lopes$^2$ \\ 
\\
$^1$ CITI / Departamento de Inform\'atica\\
Faculdade de Ci\^encias e Tecnologia,
Universidade Nova de Lisboa\\
Portugal\\
e-mail: herve@di.fct.unl.pt\\
\\
$^2$ CRACS / Departamento de Ci\^encia de Computadores \\
Faculdade de Ci\^encias,
Universidade do Porto\\
Portugal\\
e-mail: lblopes@dcc.fc.up.pt
}

\date{}

\begin{document}

\maketitle 

\begin{abstract}
Most current mobile agent systems are based on programming languages
whose semantics are difficult to prove correct as they lack an
adequate underlying formal theory. In recent years, the development of
the theory of concurrent systems, namely of process calculi, has
allowed for the first time the modeling of mobile agent systems.
Languages directly based on process calculi are, however, very
low-level and it is desirable to provide the programmer with higher
level abstractions, while keeping the semantics of the base calculus.

In this technical report we present the  syntax and the semantics of a scripting language for
programming mobile agents called \mob. \mob is service-oriented,
meaning that agents act both as servers and as clients of services and
that this coupling is done dynamically at run-time. The language is
implemented on top of a process calculus which allows us to prove that
the framework is sound by encoding its semantics into the underlying
calculus. This provides a form of language security not available to
other mobile agent languages developed using a more \emph{ah-doc}
approach.\\

\noindent
{\bf Keywords:} Mobile Computations, Service-Oriented,
  Process-Calculus, Programming Language, Run-Time System.

\end{abstract}

\section{Introduction and Motivation}
\label{sec:introduction}

The Service-oriented programming  
departs from the object-oriented  paradigm by separating the data
from the processing. Services are thus provided in a transparent way
for clients, requiring only knowledge of the contract (service's interface).
One of the main advantages of service-oriented programming is that they provide a framework on
which to develop component-based systems. In such systems,
inter-component communication is done through the contracts
provided by each component. Most of the first
service-oriented architectures were built resorting to DCOM \cite{dcom}
or to CORBA \cite{corba}. Such systems have recently
received a lot of attention for distributed systems, namely with the
.NET \cite{dotnet}, Jini \cite{jini} and Openwings \cite{openwings} platforms.

Another major technology for Web applications is that of Mobile
Agents.
Mobile agents are computations that have the ability to travel to 
multiple locations in a network, by saving their state and restoring 
it in a new host. This paradigm greatly enhances the productivity of 
each computing element in the network and creates a powerful computing 
environment, focusing on local interaction. In fact, mobile agents 
move towards the resources (e.g., data, servers) and interact locally 
unlike the usual communication paradigms (e.g., client-server), that 
require costly remote sessions to be maintained.

Programming languages for mobile agents come in two flavors: those
\emph{designed by hand} and those based on formal systems.  In the
first set we have systems such as Aglets~\cite{Lange_Oshima_1998},
Mole~\cite{mole} and Voyager~\cite{voyager} that mostly extend Java
classes to define an agent's behavior. Providing a demonstrably sound
semantics for these systems is rather difficult given the gap
between the implementation and an adequate formal model. Moreover,
since it is not possible to access the state of the Java Virtual
Machine (JVM) these systems have a hard time implementing autonomous
mobile agents, which occasionally have to move between sites carrying
their state and resuming their execution upon arrival to their
destination. Another approach, still in the same set, is that of
scripting languages such as D'Agents~\cite{d'agents} or
Ara~\cite{ara}, that fully support agent migration but require specific
virtual machine support.


Languages in the second set are based on formal systems, mostly some
form or extension of the $\pi$-calculus
\cite{async-pi:honda:tokoro:91,pi:milner:parrow:walker:92}. This
process calculus provides the theoretical framework upon which
researchers can build solid specifications for programming languages.
Languages can thus be proved correct \emph{by design} relative to some
base calculus with a well established theory. Examples of such
languages have been implemented in recent years, namely,
JoCaml~\cite{jocaml}, \ptyco~\cite{dityco:vasco:luis:fernando:98},
X-Klaim~\cite{klaim}, Nomadic
Pict~\cite{nomadic}, Acute~\cite{acute} and Alice~\cite{alice}.
Although process calculi are ideal formal tools for the development of 
mobile agent frameworks, their constructs are very low-level and 
high-level idioms that provide more intuitive abstractions for programming 
are desirable. 

Here, we introduce a scripting language called \mob that aims to
provide both language security and, a user friendly, seamless,
programming style more characteristic of the first set of languages. 
The main novelties introduced in \mob are as follows: 

\begin{itemize}
\item \mob is \emph{service oriented}, meaning that services,
  described as interfaces implemented by agents are the main 
  abstractions of the language. Agents both provide and require 
  services and they are bound to these dynamically as they move
  through the network;
  
\item the \mob language has been encoded onto a calculus that extends
  the LSD (Lexically Scoped, Distributed) $\pi$-calculus
  \cite{ravara.etal:lsd-wysiwyg} with basic objects, expressions, and
  a strong migration primitive. The $\lsdpi$-calculus is, in turn, a
  form of the $\pi$-calculus extended with support for distributed
  execution and mobility of resources and, with a well-studied
  semantics.  Although this is not the focus of this paper, we hope to
  use the encoding to prove the soundness of the operational semantics
  of the language. This is particularly important as it provides a
  form of language security, in the sense of being \emph{correct by
    design}, not readily available in related languages;

\item the encoding onto the process calculus provides a full
  specification of the front-end of the compiler for \mob. The output
  of this front-end are the \mob source programs written into
  equivalent programs in the \dityco language
  \cite{dityco:vasco:luis:fernando:98}, a concrete implementation of
  the $\lsdpi$ model. This allowed us to use both the compiler and the
  run-time system previously developed for \dityco, respectively, as
  the back-end of the \mob compiler and as the basis for the run-time
  system for \mob;

\item the \dityco run-time is implemented on top of the JVM and thus
  takes advantage of its portability while keeping full control of the
  state of the virtual machine. This provides full support for agent
  migration while keeping portability across distinct hardware
  platforms;

\item a user friendly scripting programming style provides the
  high-level abstractions desirable for programming mobile agents as
  derived constructs from the core language, thus preserving the
  semantics;

\item extensions to the core virtual machine in the form of external
  calls can be used to interact with other services (not implemented
  in \mob, namely SMTP, FTP, HTTP or SQL for databases), as well to
  execute as programs written in other programming languages, such as
  Java, Python or Perl. In this view, \mob can be used as a coordination
  language allowing high-level programming of mobile applications.
\end{itemize}

In the remainder of this paper we introduce the syntax and semantics
of the \mob programming language.









\label{sec:language}

\section{General Overview}

Before describing the language and its semantics, we give a short
overview of the main features of \mob.  We want to provide a simple to
use programming language based on the high-level type-free programming
feel of scripting languages. However, we also want to supply the means
to develop clean, modular and structured code, and therefore we adhere
to an object-oriented programming paradigm.

\emph{Agents} and \emph{services} are the two main abstractions in the
language.  Conceptually, we view an agent as a special object, with a
run-time associated, that can move from host to host in a network and
that provides/requires services to/from that network. Agents are
handled in \mob programs much like objects in object-oriented
languages: the constructor \agentk is used to define an agent
abstraction, and \newk is used to create a new instance of an agent.

Data-types are defined with the usual \classk constructor. Objects are
instances of these classes and, unlike agents, have no run-time
associated. In \mob, objects are first class entities and are also created
with the constructor \newk.

Agents may implement services (declared through the \implementsk keyword), and may,
simultaneously, be clients for the services provided by other
agents, by requiring a service (the \requiresk keyword). There is no
distinction between clients and servers.

Checking that an agent correctly uses or correctly implements the
interface of a service is done at compile time by connecting to a
network name resolver. The types inferred by the \mob compiler are
matched with those assumed for the service in the resolver. If the
agent implements a non-registered service, the interface provided
becomes the \emph{de facto} interface for that service.  This level of
type verification provides some form of program security, namely in
method invocation across agents.

To access a service, a programmer is required to get a binding for an
agent that provides it. The binding is obtained dynamically using the
\bindk primitive that asks the network resolver for an agent that
provides the required service. When the binding is received,
interaction through method invocation can happen.

Agents may move through the network and this is controlled explicitly,
at high-level by the programmer using a primitive \gok (similar to the
one found in Telescript~\cite{telescript}). The movement of an agent
involves moving an entire virtual machine and its state to the target
host in the network. The execution resumes on arrival at the target
host in a transparent way to users.

Since agents supply services, they must be able to handle multiple
incoming requests. To cope with such a demand we have designed agents
to be multi-threaded. For example, each remote method
invocation is handled by a dedicated thread. This design justified also 
the inclusion of explicit thread creation (\forkk) and synchronization 
(\joink, \waitk and \notifyk).

Objects in \mob can be accessed simultaneously by any number of
threads, therefore a scheme to allow for exclusive access is
required. We provide such a scheme with two instructions \lockk and
\unlockk that allows a primitive form of mutual exclusion in data access.

Interaction with external services is provided by \mob through the
\execk instruction. In general, \execk is used to implement extensions
to the core \mob language to support more functionality. These
external services may be implemented in other languages, such as Java,
C, TCL or Perl, or allow interaction with network services, such as
WWW queries, FTP transactions, or e-mail communication.


\begin{table}
\begin{xalignat*}{2}
\mathmob{Program} \grmeq & \mathmob{\tilde D\ P}\ ;\ \exitk & \text{A \mob program} \\
\mathmob{D} \grmeq & \mathmob{\requiresk}\ \mathmob{\tilde S}  &  \text{Require a sequence of services} \\
\grmor & \mathmob{\agentk\ X(\tilde x)\ \implementsk\ \tilde S\ \requiresk\  \tilde S\ M} & \text{Agent definition} \\
  \grmor & \mathmob{\servicek\ S\ \{ \tilde m \}} & \text{Service definition} \\
  \grmor & \mathmob{\classk\ X(\tilde x)\ M}   & \text{Class definition} \\
\mathmob{P} \grmeq &  \mathmob{I\ ;\ P} & \text{Sequential composition} \\
  \grmor & \mathmob{\epsilon} & \text{Empty sequence} \\
\mathmob{I} \grmeq & \mathmob{\gok\ (v)} & \text{Agent movement} \\
  \grmor & \mathmob{\returnk\ (v)} & \text{Method return} \\
  \grmor & \mathmob{\joink\ (x)} & \text{Thread synchronization} \\
  \grmor & \mathmob{\waitk\ (x)} & \text{Suspend on a resource} \\
  \grmor & \mathmob{\notifyk\ (x)} & \text{Wake threads suspended on a resource} \\
  \grmor & \mathmob{\lockk\ (x)} \grmor \mathmob{\unlockk\ (x)} & \text{Lock/unlock resource} \\
  \grmor & \mathmob{\ifk\ (v)\ \{\  P\ \}\  \elsek\ \{\  P\ \}} & \text{Conditional execution} \\
  \grmor & \mathmob{\whilek\ (v)\ \{\  P\ \}} & \text{Iterator} \\
  \grmor & \mathmob{\breakk}\ & \text{Break} \\
  \grmor & \mathmob{\exitk} & \text{Terminate agent execution} \\
  \grmor & \mathmob{x = V} & \text{Assignment} \\ 
  \grmor & \mathmob{o.x = v} & \text{Attribute assignment} \\     
\mathmob{V} \grmeq & \mathmob{\newk\ X\ (\tilde v)} & \text{New agent or object} \\
  \grmor  & \mathmob{\forkk\ \{\ P\ \}} & \text{New thread} \\
  \grmor & \mathmob{\bindk\ (S\ v) \grmor \bindk(S)} & \text{Service discovery} \\
  \grmor & \mathmob{\hostk\ ()} & \text{Current host} \\
  \grmor & \mathmob{\execk\ (\tilde v)} & \text{External service call} \\
  \grmor & \mathmob{o.m\ (\tilde v)} & \text{Method invocation} \\
  \grmor  & \mathmob{e} & \text{Expressions} \\
  \grmor  & \mathmob{o.x} & \text{Attribute reading} \\
\mathmob{M} \grmeq & \mathmob{\{ m_1 ( \tilde{x}_1 )\ \{\  P_1\ \} \dots m_n ( \tilde{x}_n )\ \{\ P_n\ \} \}}
	& \text{Methods} \\
\mathmob{e}  \grmeq & \mathmob{v  \grmor\ e\ bop\ e \grmor\ uop\ e} & \text{Basic expressions} \\
\mathmob{v}  \grmeq & \mathmob{o \grmor\ c \grmor \nullk}   & \text{Language values} \\
\mathmob{o}  \grmeq & \mathmob{x \grmor \selfk}  & \text{Target} \\
\mathmob{c}  \grmeq & \mathmob{bool \grmor int \grmor string}  & \text{Constants}\\
\mathmob{bop}  \grmeq &  + \grmor -  \grmor *  \grmor / \grmor \% \grmor \hat{} \\ 
\grmor  & \&\& \grmor || \grmor  == \grmor != \\
\grmor  & < \grmor > \grmor  <= \grmor >=   & \text{Binary Operators} \\
\mathmob{uop}  \grmeq & ! \grmor - & \text{Unary Operators} 
\end{xalignat*} 
\caption{Syntax of the \mob programming language}
\label{fig:grammar}
\end{table}

\section{The Syntax}
\label{sec:lang:syntax}

We now present the syntax for the core language. The full form of the
language is obtained by providing derived constructs for higher-level
programming, while keeping the underlying semantics of the core
language.

As defined in the grammar in table \ref{fig:grammar}, a \mob program is syntactically a
sequence of definitions ($\mathmob{\tilde D}$) followed by a sequence
of instructions ($\mathmob{P}$) terminated by \exitk. %
Tables \ref{fig:cats} and \ref{fig:reserved} present, respectively, the 
phrase categories and the set of reserved words, that may not be used as identifiers, required by the  syntax of a \mob program.

We choose only to allow the assignment of language values to class and
agent attributes. The assignment of an element in $\mathmob{V}$ to an
attribute must use an intermediate auxiliary variable. This will be
overcame in the full form of the language, with the inclusion of
syntactic sugar. Here, it greatly simplifies the definition of the
language's semantics, since we do not have to define duplicate rules
for the assignment of the elements in $\mathmob{V}$ to either
variables and attributes.

Constant identifiers in \mob are divided in the following classes:
booleans, elements of \linebreak \Bool = \{\truek, \falsek\} ranged
over by $\mathmob{bool}$; integers, elements of \Int ranged over by
$\mathmob{int}$ and, strings, elements of \String ranged over by
$\mathmob{string}$, defined by the regular expression: $"[\hat\
\backslash" \backslash n]^*"$.  Hosts in \mob are represented with strings.

\begin{table}
\begin{center}
\begin{tabular}{lcll}
  $\mathmob {Program}$  & $\in$ & \text{Program} & \text{A \mob program} \\
  $\mathmob D$  & $\in$ & \text{Definition} & \text{Class, agent and
    service definitions,}\\
&&& \text{and service requirement} \\
$\mathmob{X,Y}$  & $\in$ & \Class & \text{Class or agent identifiers}\\
$\mathmob{S}$  & $\in$ & \ServiceId & \text{Service identifiers} \\
$\mathmob{m}$    & $\in$ & \MethodId & \text{Method labels} \\
$\mathmob{P}$  & $\in$ & \text{InstructionSeq} \quad & \text{Sequence of program instructions} \\
$\mathmob{I}$  & $\in$ & \text{Instruction} & \text{Program instruction} \\
$\mathmob{V}$  & $\in$ & \text{AssignValue} & \text{Assignable value}\\
$\mathmob M$  & $\in$ & \text{Method} & \text{Set of methods} \\
$\mathmob e$  & $\in$ & \text{Expression} & \text{Basic expression}\\
$\mathmob{x,y}$  & $\in$ & \Var & \text{Variable identifier}\\
$\mathmob c$  & $\in$ & \text{Constant} & \text{Constant value}\\
$\mathmob v$  & $\in$ & \text{LangValue} & \text{Variable or constant} \\
$\mathmob o$  & $\in$ & \text{Target} & \text{Object or agent} \\
$\mathmob{bop}$  & $\in$ & \text{BinOp} & \text{Binary operations}\\
$\mathmob{uop}$ & $\in$ & \text{UnOp} & \text{Unary operations}
\end{tabular}
\end{center}
\caption{Phrase categories}
\label{fig:cats}
\end{table}

\begin{table}
\begin{gather*}
  \agentk \qquad  \implementsk  \qquad \requiresk \qquad \classk
  \qquad \servicek  \qquad \maink \qquad \newk \qquad \\
  \gok \qquad \bindk \qquad \forkk \qquad  \joink \qquad \waitk \qquad  \notifyk \qquad
  \lockk \qquad \unlockk \qquad   \hostk \qquad \execk  \\
  \ifk \qquad \elsek \qquad \whilek \qquad \breakk \qquad \returnk
  \qquad \exitk \qquad  \selfk \qquad \nullk \\
! \qquad \{ \qquad \} \qquad ( \qquad ) \qquad  . \qquad ; \\
+ \qquad - \qquad * \qquad / \qquad \% \qquad
== \qquad != \qquad > \qquad < \qquad >= \qquad <= \qquad \&\& \qquad ||
\end{gather*}
\caption{Reserved words}
\label{fig:reserved}
\end{table}

\subsubsection*{Syntactic Restrictions}
\label{sec:lang:syntax-res}

The concrete syntax of \mob imposes some syntactic restrictions over
the syntax of a program:

\begin{itemize}
\item service definitions must precede service requirements that, in
  turn, must precede the definitions of classes and agents;

\item an agent must implement the \maink method;

\item the \returnk instruction can only appear within the body of a method;

\item the \breakk instruction can only appear inside the body of a \whilek instruction;

\item the \gok and \exitk instructions can only appear inside the body
  of a method in an  agent definition;

\item the method identifiers $\mathmob{\tilde m}$ in $\mathmob{\{ m_1 ( \tilde{x}_1 )\ \{\
P_1\ \} \dots m_n ( \tilde{x}_n )\ \{\ P_n\ \} \}}$, and in $\servicek\ \{ \mathmob{\tilde m}\}$ are
pairwise distinct;

\item the parameters $\mathmob{\tilde x}$ in an agent ($\mathmob{\agentk\ X(\tilde x)\
\implementsk\ \tilde S\ \requiresk\ \tilde S\ M}$), a class ($\mathmob{\classk\ X(\tilde x)\ M}$) or a
method ($\mathmob{m(\tilde x)\{ P\}}$) definitions are pairwise distinct;

\item a variable $\mathmob{x}$ is bound in $\mathmob{P}$ with an assignment ($\mathmob{x=V;P}$),
  where $\mathmob{V}$ is a language construct that may appear on the left of an
  assignment; is bound in $\mathmob{M}$ in an agent definition \linebreak($\mathmob{\agentk\
  X(\tilde x)\ \implementsk\ \tilde S\ \requiresk\ \tilde S\ M}$) or
  class definition ($\mathmob{\classk\ X(\tilde x)\ M}$) if it is one of the
  $\mathmob{\tilde x}$; is bound in $\mathmob{P}$ in a method ($\mathmob{m(\tilde x)\{P\}}$) if it
  is one of the $\mathmob{\tilde x}$;

\item an agent identifier $\mathmob{X}$ is bound in $\mathmob{\tilde  D}$,  $\mathmob{\tilde  D'}$, $\mathmob{M}$ and 
  $\mathmob{P}$ with a statement \linebreak $\mathmob{\tilde D\ \agentk\ X(\tilde x)\ \implementsk\ \tilde S\ \requiresk\ \tilde
    S\ M\ \tilde D'\ P}$;

\item a class identifier $\mathmob{X}$ is bound in  $\mathmob{\tilde  D}$,  $\mathmob{\tilde  D'}$, $\mathmob{M}$ and 
  $\mathmob{P}$  with a 
statement $\mathmob{\tilde D\ \classk\ X(\tilde x)\ M\ \tilde D'\ P}$;

\item a service identifier $\mathmob{S}$ is bound in
  $\mathmob{\tilde  D}$  and 
  $\mathmob{P}$ with a statement 
$\mathmob{\servicek\ S\ \{ \tilde m\}\ \tilde D\ P}$;

\item the sets of free variables, free agents, free classes and free 
services are defined accordingly. Well formed \mob{} programs are 
closed for variables, agent identifiers, class identifiers and 
service identifiers. 
\end{itemize}


\section{The \mob Abstract Machine}
\label{sec:mobam}

We provide the semantics for a \mob network in the form of an abstract state
transition machine.  A \mob network is composed by a set of hosts,
which are abstractions for network nodes. Hosts define the boundaries
where computations take place in a \mob network. The computing units
of the \mob language are agents. There may be several agents running
concurrently in a given host at any given time. In our approach there
is no distinction between clients and servers, any agent may behave as
a client requesting a service while also providing services to
others. This is achieved by implementing multi-threaded agents to
handle multiple requests concurrently.  The threads in an agent share
the same heap space whilst having independent control data-structures.

Before defining the structure of the network, of agents and of threads we
first introduce the syntactic categories and auxiliary
functions.

\subsection{Syntactic Categories and Data-Structures}

Besides the syntactic categories of the language, the abstract machine
requires a new set of categories\footnote{We denote the syntactic
  definition of methods in the language and their internal
  representation in the abstract machine with the same letter
  $\mathmob{M}$. We choose to do so because both relate to the same
  information, although with different representations.} defined in table \ref{fig:vmcats}.

\begin{table}
\begin{center}
\begin{tabular}{lcll}
  $\mathmob{a, b}$  & $\in$ & \text{AgentKey} & \text{Agent key} \\
  $\mathmob{C}$ & $\in$ & \text{Code} & \text{Code for the class and  agent definitions} \\ 
  $\mathmob{M}$ & $\in$ &  \text{Method} & \text{Methods of a
    definition} \\ 
  $\mathmob{H}$ &   $\in$ & \text{Heap} & \text{Address space for the agent} \\
  $\mathmob{r}$ & $\in$ & \text{HeapRef} & \text{Heap reference} \\
  $\mathmob{t}$ & $\in$ & \text{ThreadRef} & \text{Thread reference}
  \\ 
  $\mathmob{K}$ & $\in$ & \text{Closure} & \text{Closure} \\
  $\mathmob{u}$ & $\in$ & \text{Value} & \text{Constant value or a
    heap reference} \\ 
  $\mathmob{T}$ & $\in$ & \text{Pool(RunningThread)} \quad & \text{Pool of flows of execution} \\
  $\mathmob{B}$ & $\in$ & \Environment & \text{Environment of a
    thread} \\ 
  $\mathmob{Q}$ & $\in$ & \text{CodeStack} & \text{Stack of interrupted blocks of code} \\ 
  $\mathmob{W}$ & $\in$ & \text{SuspendedThread} & \text{Threads suspended (waiting) on a heap
  reference} \\ 
$\calA$ & $\in$ & \text{Pool(Agent)} & \text{Pool of
  agents}\\ 
$\alpha$ & $\in$ & \text{Type} & \text{Type of a service}
  \\ 
  $\ns$ & $\in$ & \NS & \text{Name resolver} \\ 
  $\mathmob{ANS}$ & $\in$ & \ANS & \text{Agent name resolver} 
  \\ $\mathmob{SNS}$ & $\in$ & \SNS & \text{Service name resolver} \\ 
  $\calN$ & $\in$ &  \text{Network} & \text{Network}
\end{tabular}
\end{center}
\caption{Syntactic categories of the \mob virtual machine}
\label{fig:vmcats} 
\end{table}


The abstract machine has two layers: agents and threads. A
network is described as a set of agents running concurrently plus a
resolver for agents and services. Agents are described as collections of threads 
running concurrently and sharing the agent's resources, namely its code and 
address space.
Agents are abstractions for autonomous programs running on network
hosts, that interact with the network by spawning new agents or moving
between hosts. Inter-agent interaction is performed by invoking methods.
The abstract machine requires some syntactic categories and
data-structures to be defined:

\begin{itemize}

\item an \emph{Agent Key} is an element of the set $\AgentName \subset
  \String$, ranged over by $\mathmob{a, b}$, and represents a unique,
  network-wide, key for an agent;

\item a \emph{Host} is an element of the set $\Host \subset \String$,
  ranged over by $\mathmob{h}$, and represents a unique, network-wide,
  host identifier;

\item an \emph{Instruction} is an element of the set $\Instruction$,
  ranged over by $\mathmob{I}$, and represents a \mob instruction. A
  sequence of instructions separated by \textbf{;} is denoted by $\mathmob{P} \in \text{InstructionSeq}$;

\item \emph{Method} represents a set of methods and is a map of the
  form $\Method = \MethodId \mapsto \Var^* \times \InstructionSeq$,
  ranged over by $\mathmob{M}$, and represents the methods in a class
  or an agent;

\item the \emph{Code} repository for an agent is a map defined as
  $\Code = \Class \mapsto \Bool \times  \Var^* \times
  \Method \times \Code \times \ServiceId^*$, ranged over by $\mathmob{C}$ and
  represents the all the code required by a class or agent.  The
  boolean value makes the distinction between the two (\truek\ = agent,
  \falsek\ = class);

\item a \emph{Heap Reference} is an element of the set $\HeapRef$,
  ranged over by $\mathmob{r}$, and is an abstraction for an address
  in the address space of an agent.
  Heap references in \mob are qualified with the key of their hosting
  agent (e.g., reference $\mathmob{r}$ in the heap of agent
  $\mathmob{a}$ should be interpreted as $\mathmob{r@a}$) and thus are
  unique in the network. To ease the reading of the rules we omit the
  qualifier of a heap reference when it is accessed from within its
  hosting agent. The value $\nullr \in \HeapRef$ represents an
  undefined heap reference;

\item a \emph{Thread Reference} is an element of the $\ThreadRef
  \subset \HeapRef$, ranged over by $\mathmob{t}$, and represents a
  reference to a thread. Note that this subset includes $\nullr$;

\item a \emph{Constant} is an element of the set $\Constant = \Bool
  \cup \Int \cup \String$, ranged over by $\mathmob{c}$, and represents a
  primitive value of the language;

\item a \emph{Value} is an element of the set $\Value = \Constant \cup
  \HeapRef$, ranged over by $\mathmob{u}$;

\item an \emph{Environment} is a map defined as $\Environment = \Var
  \mapsto \Value$, ranged over by $\mathmob{B}$, that represents a map from
  identifiers in the code to constants or references in the heap. We will represent
  the binding from an identifier $\mathmob{x}$ to a value $\mathmob{u}$ as $\mathmob{x:u}$;

\item a \emph{Closure} is an element of the set $\Closure = \Bool
  \times \Environment \times \Class$, ranged over by $\mathmob{K}$,
  and represents the closure for an instance of a class or an agent
  located in an address space. The boolean value makes the distinction
  between instances of classes and agents;

\item a \emph{Heap} is a map defined as $\Heap = \HeapRef \mapsto
  \ThreadRef \times (\Closure \cup \Value)$, \linebreak ranged over by
  $\mathmob{H}$, and represents the address space of an agent. The contents
  associated with heap references may be accessed with mutual
  exclusion using locks.  The thread reference
  in the image of a reference indicates which thread holds the lock
  to the contents. Unlocked references hold  $\nullr$ has their
  thread reference;

\item a \emph{Code Stack} is an element of the set $\Stack =
  Stack(\Environment \times \InstructionSeq)$, ranged over by $\mathmob{Q}$, and
  represents the stack of blocks of code, used as a mechanism to
  implement \whilek loops. The block of the top of the stack is the
  one currently being executed by the machine. As soon as it
  terminates, it is popped from the stack and the execution continues
  with the code of the block found at the top of the stack.  We refer to elements
  of the stack as \emph{code-blocks};

\item a \emph{Pool of Running Threads} is an element of the set
  $\Pool(\RunningThread)$, ranged over by $\mathmob{T}$, where
  $\RunningThread = \ThreadRef \times \Stack \times \HeapRef$ 
  represents a flow of execution. In the definition of a thread
  $\mathmob{(t, Q, r)}$, the thread reference $\mathmob{t}$ is a
  location in the heap to which the thread is bound. The reference
  $\mathmob{r}$ is a heap reference where the thread may place a
  result;

\item a \emph{Suspended Thread} is a map defined as $\SuspendedThread
  = \HeapRef \mapsto 2^\RunningThread$, ranged over by $\mathmob{W}$, and
  represents threads suspended (waiting) on heap references;

\item a \emph{Pool of Agents} is an element of the set
  $\Pool(\Agent)$, ranged over by $\calA$, where $\Agent = \AgentName
  \times \Host \times \Code \times \Heap \times \Pool(\RunningThread)
  \times \SuspendedThread$ represents a multi-threaded autonomous
  computation. We write an agent $\mathmob{(a,h,C,H,T,W)}$ as
  $\mathmob{a(h,C,H,T,W)}$ thus exposing the agent's key;
 
\item a \emph{Service Type} is an element of the set $\Type$, ranged
  over by $\alpha$;

\item a \emph{Name Resolver}, $\calR$, is composed by two maps, $\NS = \ANS \times$ \linebreak
  $\SNS$. The first, defined as $\ANS = \HeapRef \mapsto \Host$, ranged
  over by $\mathmob{ANS}$, represents a network-wide name resolver for locating
  agents. The second, defined as $\SNS = \ServiceId \mapsto \Type
  \times 2^\HeapRef$, ranged over by $\mathmob{SNS}$, represents a network-wide
  name resolver for obtaining the type and implementations of a service;


\item a \emph{Network} is an element of the set $\Network =
  \Pool(\Agent) \times \NS$, ranged over by $\calN$, and
  represents a \mob network computation.

\end{itemize}

\subsection{Auxiliary Definitions}
Function \access checks if, in a heap $\mathmob{H}$, the access to the value
located at $\mathmob{r}$ is granted to a thread identified by $\mathmob{t}$.

\begin{step}
\begin{array}{rlll}
\access: \Heap \times \HeapRef \times \ThreadRef & \mapsto & \Bool \\
\access(\mathmob{H, r, t}) & = &
 \left \{ 
 \begin{array}{ll}
   \truek, & \text{if}\ \mathmob{H(r) = (t, \_)} \;\text{or}\; \mathmob{H(r) = (\nullr, \_)}\\
   \falsek, & \text{if}\ \mathmob{H(r) = (t', \_)} \;\text{and}\; \mathmob{t' \not = t}
 \end{array} \right. 
\end{array}
\end{step}

Function \locked tries to grant the lock for a value located at
$\mathmob{r}$, in a heap $\mathmob{H}$,
to a thread identified by $\mathmob{t}$. Function \unlocked tries
to release the lock for a value located at $\mathmob{r}$ in a heap $\mathmob{H}$. The
result of both functions is a heap modified (or not) by the operation,
and a boolean value indicating if the operation was successful. Both
these functions are atomic.

\begin{step}
\begin{array}{rlll}
 \locked: \Heap \times \HeapRef \times \ThreadRef & \mapsto & \Heap \times \Bool \\
 \locked(\mathmob{H, r, t}) & = & 
 \left \{ 
 \begin{array}{ll}
   \mathmob{(H + \{r : (t, K)\}, \truek)}, & \text{if}\ \mathmob{\access(H, r, t) =
   \truek}\\ & \text{and}\ \mathmob{H(r) = (\_, K)}\\
   \mathmob{(H + \{r : (t, u)\}, \truek)}, & \text{if}\ \mathmob{\access(H, r, t) =
   \truek}\\ & \text{and}\ \mathmob{H(r) = (\_, u)}\\
   \mathmob{(H, \falsek)}, & \text{if}\  \mathmob{\access(H, r, t) = \falsek}
 \end{array} \right. 
\end{array}
\end{step}

\begin{step}
\begin{array}{l}
 \unlocked: \Heap \times 
\HeapRef  \times \ThreadRef  \mapsto  \Heap \times \Bool \\
 \hspace{3.5cm} \unlocked(\mathmob{H, r, t})   = 
 \left \{ 
 \begin{array}{ll}
   \mathmob{(H + \{r : (\nullr, K)\}, \truek)}, & \text{if}\ \mathmob{\access(H, r, t) =
   \truek}\  \\ & \text{and}\ \mathmob{H(r) = (\_, K)}\\
   \mathmob{(H + \{r : (\nullr, u)\}, \truek)}, & \text{if}\ \mathmob{\access(H, r, t) =
   \truek}\\  & \text{and}\ \mathmob{H(r) = (\_, u)}\\
   \mathmob{(H, \falsek)}, & \text{if}\  \mathmob{\access(H, r, t) = \falsek}
 \end{array} \right. 
\end{array}
\end{step}

Remember that unlocked references have $\nullr$ as their locking reference.

Function \codeOf returns the code for a method $\mathmob{m}$ from a
given class or agent representation:

\begin{step}
\begin{array}{rlll}
 \codeOf: (\Var^* \times \Method \times \ServiceId^*) \times \MethodId
 & \mapsto & \Var^* \times \InstructionSeq \\
 \mathmob{\codeOf((\tilde x, M, \tilde S), m)} & = & \mathmob{M(m)}
\end{array}
\end{step}

Function \codein returns the code closure for a set of methods. The
result is a code repository, built from another received as argument,
that is composed of the code for all the classes referenced in the set of methods.

\begin{step}
\begin{array}{rlll}
\codein:  \Code \times \Method & \mapsto & \Code\\
 \mathmob{\codein(C, M \cdot (\tilde x, x = \newk\ X(\tilde v); P))} & = &  \mathmob{\{X : C(X)\} +\codein(C, M \cdot (\tilde x, P))}\\
 \mathmob{\codein(C, M \cdot (\tilde x, I; P))} & = & \mathmob{\codein(C, M \cdot (\tilde x, P))} & \text{if}\ \mathmob{I \not = x = \newk\ X(\tilde v);}\\
 \mathmob{\codein(C, M \cdot (\tilde x, \epsilon))} & = & \mathmob{\codein(C, M)}\\
 \mathmob{\codein(\emptyset)} & = & \emptyset
\end{array}
\end{step}

Function \evalseq returns the evaluation of a sequence of expressions,
each element being evaluated by the \evalone function.  The evaluation
requires the knowledge of the state of the heap $\mathmob{H}$, the heap
reference of the thread computing the expression $\mathmob{r}$, and its
environment $\mathmob{B}$.  The evaluation is fairly standard, however a particularity requires some closer attention.
A variable may contain a heap reference whose value is another
reference, which forces the resolution of the indirection.


The value of a constant is given by the built-in \textsf{val}
function, and the result of the relational and arithmetic built-in
operations is given by the \textsf{bop} and
\textsf{uop} built-in functions.

\begin{step}
 \evalseq : (\Heap \times \ThreadRef \times \Environment \times
\Expression^*) & \mapsto & \Value^* \\
 \mathmob{\evalseq(H, t, B, v\ \tilde v)} & = & \mathmob{\evalone(H, t, B, v)\ \evalseq(H, t, B, \tilde v)} \\
 \mathmob{\evalseq(H, t, B, \epsilon)} & = & \mathmob{\epsilon}
\end{step}

\begin{step}
\begin{array}{rlll}
 \evalone : (\Heap \times \ThreadRef \times \Environment \times
 \Expression) & \mapsto & \Value\\
 \mathmob{\evalone(H, t, B, c)} & = & \textsf{val}(\mathmob{c}) \\
 \mathmob{\evalone(H, t, B, \nullk)} & = & \nullr\\
 \mathmob{\evalone(H, t, B, x)} & = & \textsf{val}(\mathmob{c}) & \text{if}\ \mathmob{B(x) = c} \\ 
 \mathmob{\evalone(H, t, B, x)} & = & \nullr & \text{if}\ \mathmob{B(x) = \nullr}\\
 \mathmob{\evalone(H, t, B, x)} & = & \mathmob{t'} & \text{if}\ \mathmob{B(x) = t'} \\
 \mathmob{\evalone(H, t, B, x)} & = & \mathmob{r} & \text{if}\ \mathmob{B(x) = r}\ \text{and}\ \mathmob{H(r) = (\_, K)} \\
 \mathmob{\evalone(H, t, B, x)} & = & \mathmob{u} & \text{if}\  \mathmob{B(x) = r}\ \text{and}\ \mathmob{H(r) = (\_, u)} \\
 \mathmob{\evalone(H, t, B, e\ bop\ e')} & = & \textsf{bop}(\mathmob{u, u'})  \qquad& \text{if}\
 \mathmob{\evalone(H, t, B, e) = u}\\
&&& \text{and}\ \mathmob{\evalone(H, t, B, e') = u'} \\
 \mathmob{\evalone(H, t, B, uop\ e)} & = & \textsf{uop}(\mathmob{u}) & \text{if}\    \mathmob{\evalone(H, t, B, e) = u}\ 
\end{array}
\end{step}

Function $\mathmob{\copyseq_{ab}}$ returns a copy of the closures for a sequence
of values located in the heap of an agent $\mathmob{a}$, plus all the code they
require.  The new references created to duplicate the given closure
are located in the target agent $\mathmob{b}$.  The function takes as arguments
the code repository $\mathmob{C}$ and heap $\mathmob{H}$ of the original agent $\mathmob{a}$, and the
sequence of values to be copied $\mathmob{\tilde u}$. The copy of each value is
computed by function $\mathmob{\copyone_{ab}}$.

\begin{step}
\begin{array}{rlll}
 \mathmob{\copyseq_{ab}}: \Code \times \Heap \times \Value^* & \mapsto & \Code \times \Heap \times \Value^* \\
\mathmob{\copyseq_{ab}(C, H, u\ \tilde u)} & = & \mathmob{(C' + C'', H' + H'', u' \tilde
u')} & \text{where}\ \mathmob{\copyone_{ab}(C, H, u) = (C', H', u')}\\
&&& \text{and}\ \mathmob{\copyseq_{ab}(C, H, \tilde u) = (C'', H'', \tilde u')} \\
\mathmob{\copyseq_{ab}(C, H, \epsilon)} & = & \mathmob{(\emptyset, \emptyset, \epsilon)}
\end{array}
\end{step}

\begin{step}
\begin{array}{rlll}
 \mathmob{\copyone_{ab}}: \Code \times \Heap \times \Value & \mapsto & \Code \times \Heap \times \Value \\
 \mathmob{\copyone_{ab}(C, H, r@a)}  & = &  \mathmob{(\{X : C(X)\} + C', H' + \{r'@b :
 (\nullr, (\falsek, \{\tilde x : \tilde u'\}, X)\}, r'@b))}\\
\end{array}
\end{step}

\begin{step}
\hspace{1.9cm} 
\begin{array}{rlll}
&&&  \text{if}\ \mathmob{H(r@a) = (\_, (\falsek, \{\tilde x : \tilde u\}, X))}\\ 
&&&   \text{where}\ \mathmob{\copyseq_{ab}(C, H, \tilde u) = (C',  H',
  \tilde u')} \\ 
&&& \text{and}\  \mathmob{r'@b} \in \HeapRef\  \fresh \\
 \mathmob{\copyone_{ab}(C, H, r@a)}  & = &  \mathmob{(\emptyset, \emptyset, r@a)}\ &
      \text{if}\ \mathmob{H(r@a) = (\_, (\truek, \{\tilde x : \tilde  u\}, X))}  \\
  \mathmob{\copyone_{ab}(C, H, r@a)}  & = & \mathmob{(C', H' + \{r'@b : u'\}, r'@b)} &   \text{if}\ \mathmob{H(r@a) = (\_, u)}\\
&&&         \text{where}\ \mathmob{\copyone_{ab}(C, H, u) = (C', H', u')}\\ 
&&&\text{and}\  \mathmob{r'@b} \in \HeapRef\  \fresh  \\
  \mathmob{\copyone_{ab}(C, H, r@a')} & = & \mathmob{(\emptyset,
    \emptyset, r@a')}\ &   \text{if}\ \mathmob{a' \not = a \not = b} \\
  \mathmob{\copyone_{ab}(C, H, c)} & = & \mathmob{(\emptyset, \emptyset, c)}
\end{array}
\end{step}

The \run function places a set of pool of running threads in concurrent
execution.

\begin{step}
\begin{array}{rlll}
\run:  2^{\RunningThread} & \mapsto & \Pool(\RunningThread)\\
 \mathmob{\run(\{(t_1, Q_1, r_1), \dots, (t_n, Q_n, r_n)\})} & =&  \mathmob{(t_1, Q_1, r_1) \conc \cdots \conc (t_n, Q_n, r_n)}
\end{array}
\end{step}

\subsection{The Initial and Final States}

Based on the above definitions, we may write the syntax for a network
as follows: 

\begin{xalignat*}{2}
\mathmob{\calN} \grmeq & \mathmob{\calA, \ns} & \text{Network} \\         
\mathmob{\calA} \grmeq & \mathmob{\calA \conc \calA} & \text{Concurrent agents} \\
                \grmor & \mathmob{a(h, C,H,T,W)} & \text{Running agent} \\
                \grmor & \mathmob{\zeroka} & \text{Terminated agent} \\
\mathmob{T}     \grmeq & \mathmob{T \conc T} & \text{Concurrent threads} \\
                \grmor & \mathmob{(t, Q, r)} & \text{Running thread}\\
                \grmor & \mathmob{\zerokt} & \text{Terminated thread} 
\end{xalignat*} 

For the sake of simplicity we assume that agents run in a static
network with no failures. In other words, the set of available hosts, \Host, 
is constant.  Here we describe the
abstract machine from the point of view of the execution of one
agent. Thus, when we start running an agent, the network may already
have a pool of agents $\calA$ running concurrently and distributed among
the network nodes in the set $\Host$, together with the resolver $\ns$:

         \begin{equation*}
                \calA , \ns
         \end{equation*}

We launch a program ($\mathmob{\tilde D\ P}$) in the network by encapsulating its code
($\mathmob{P}$) in
an agent that is placed in a host specified by the user. 
%
The code repository for the program is collected at compile-time with
function $\codeclo$ that we will defined ahead. Thus, when the agent
is launched into the network it already contains all the code it
requires.  The initial state of the execution of a program with code
$\mathmob{\tilde D\ P}$ is thus:

\begin{step}
  \mathmob{a(h, \codeclo(\tilde D), \emptyset, \launchT(\emptyset, P, \nullr),\emptyset) \conc \calA, \ns}
\end{step}

\noindent
where $\mathmob{a}$ is a fresh agent key, $\mathmob{h}$ is the local
host and $\mathmob{\launchT(B,P,r)}$ is a macro that creates a new
thread with an environment $\mathmob{B}$ (here $\emptyset$), a code
$\mathmob{P}$ (here $\mathmob{P}$) and a return reference
$\mathmob{r}$ (here $\nullr$):

\begin{step}
  \mathmob{a(h, C, H, \launchT(B, P, r) \conc T, W) \conc \calA, \calR}  & \eqdef &\\
  \mathmob{a(h, C, H + \{t : (t, \nullr)\}, (t, (B, P), r) \conc T, W)  \conc \calA, \calR} && \qquad \mathmob{t} \in 
\ThreadRef\ \fresh
\end{step}

Note that no heap reference is associated with the agent $\mathmob{a}$, since a
program does not provide any methods, nor has attributes. Moreover,
$\mathmob{a}$ is not registered in $\ns$, and thus is not accessible to the
network.
The registry is a precondition for an agent to migrate (further detail 
in rule \rulename{Go}), and thus, \mob programs cannot migrate, just agents.

Agents are daemons by default and must be explicitly terminated by the
\exitk instruction, which produces the terminated agent
$\mathmob{\zeroka}$. Thus, at the end of the program running in agent
$\mathmob{a}$, the configuration of the network will be of the form:

         \begin{step}
                \zeroka \conc \calA', \ns'
         \end{step}

Such an agent can thus be garbage collected and produce the state:

         \begin{step}
           \calA', \ns'
         \end{step}

\subsection{Code Collection}

The compile-time code collection is defined by function $\codeclo:  \Def^*  \mapsto  \Code$ that
returns a code repository with all the code required  by a sequence of class and
agent definitions. We present the function in a case by case analysis.

A service specifies an interface implemented by some \mob agent.
Service definitions are used to supply information to the type-system.
Type-checking of a \mob program is performed at compile-time by
matching the inferred types for services required or implemented by
the agents with their definitions kept in the resolver. If the service
is required by the program or if the program implements a known
service in the network then its inferred type must match the interface
for the service kept in the resolver. If the service is introduced for
the first time by the program (an interface for it does not yet exist
in the resolver) then the type inferred for the service will become
the adopted interface for the service as registered in the resolver.
So, when the agent is created, the $\mathmob{SNS}$ map is updated by
adding the reference of the agent to every entry associated to a
one of the implemented services.
Anyway, these are handled at compile time and there is no need for
them in the abstract machine.

\begin{step}
 \mathmob{\codeclo(\servicek\ S\ \{ \tilde m\}\ \tilde D)}  & = & \mathmob{\codeclo(\tilde D)}
\end{step}

Simple classes define abstract data-types and we call their instances
\emph{objects}. The entry in the code repository associated with this
definition contains a closure with slots for the code for all the
classes and agents that are required by this class, its attributes,
the code for its methods, and an empty sequence of implemented
services, since an object does not provide any services. Remember that
\falsek\ indicates that the entry contains the code of a class.

\begin{step}
\begin{array}{rlll}
 \mathmob{\codeclo(\classk\ X(\tilde x)\  \{  m_1 ( \tilde{x}_1 )\ \{\ P_1\ \} \dots m_n ( \tilde{x}_n
)\ \{\ P_n\ \} \}\ \tilde D)}  & = & \\
  \mathmob{\{X: (\falsek, \tilde{x}, M', \codein(M'), \epsilon) \}  +
    \codeclo(\tilde D)}
\end{array}
\end{step}

\noindent
where $\mathmob{M' = \{m_1 : (\tilde{x}_1,P_1), \dots, m_n : (\tilde{x}_n,P_n)\}}$.

Some classes are special in the fact that they represent full computations.  We
call their instances \emph{agents}, and we use a different
keyword to differentiate them. Otherwise, the definition of an agent is very much
like that of a regular class, it contains the code for all the
classes and agents required, the attributes, the code for the methods
and indicates which services are provided by the agent.

The \requiresk keyword supplies information to the type-system,
indicating which services are required by the agent and that their
uses must be checked against the definitions in the $\mathmob{SNS}$.
The \implementsk keyword does not only supply information to the
type-system, but also states which service entries must be updated
whenever an instance of the agent is created (rule
\rulename{NewAgent}).  To hold this information when necessary, we
keep this sequence in the agent's code closure.

\begin{step}
\begin{array}{rlll}
 \mathmob{\codeclo(\agentk\ X(\tilde x)\ \implementsk\  \tilde S\
   \requiresk\ \tilde S'\ \{ m_1 ( \tilde{x}_1 )\ \{\ P_1\ \} \dots m_n ( \tilde{x}_n
)\ \{\ P_n\ \} \}\  \tilde D)}  & = & \\
 \mathmob{\{X: (\truek, \tilde{x}, M', \codein(M'), \tilde S) \} +
   \codeclo(\tilde D)}
\end{array}
\end{step}

\noindent
where $\mathmob{M' = \{m_1 : (\tilde{x}_1,P_1), \dots, m_n : (\tilde{x}_n,P_n)\}}$.

Note that agents may not always implement or require services and thus
both $\mathmob{\tilde S}$ or $\mathmob{\tilde S'}$ may be empty
sequences.

The \requiresk\ keyword can be also used by itself to indicate which are
the services required by a program. Once again this only provides
information to the compile time type checking, and thus there is no
need to pass it to the run-time. Here we also present the base case
for the recursion.

\begin{step}
\begin{array}{rlll}
 \mathmob{\codeclo(\requiresk\ \tilde S\ \tilde D)} & = & \mathmob{\codeclo(\tilde D)}\\
 \mathmob{\codeclo(\epsilon)} & = & \emptyset
\end{array}
\end{step}

\subsection{The Congruence Rules}
The computation in the abstract machine is driven by a set of
 \emph{reduction rules} that operate over the thread or the agent at
 the most left in the respective pool.
Thus, in order to be able to commute, associate and garbage collect threads and
agents in their pools, we need a set of \emph{congruence rules}. These
will allow for the re-writing of both pools, into semantically
equivalent ones, where the configuration is  accordingly to the
reduction rules to be applied.
The congruence rules for a pool of agents are:

\begin{multicols}{2}
\crule{
  \rulename{AgentSwap}
}{
  \calA \conc \calA'
}{
  \calA' \conc \calA
} 

\crule{
  \rulename{AgentAssoc}
}{
  \calA \conc (\calA' \conc \calA'')
}{
  (\calA \conc \calA') \conc \calA''
}

\crule{
  \rulename{AgentGC}
}{
  \zeroka \conc \calA
}{
  \calA
}


\end{multicols}
\bigskip


The congruence rules for threads for threads are:

\begin{multicols}{2}
\crule{
  \rulename{ThreadSwap}
}{
  T \conc T'
}{
  T' \conc T
} 

\crule{
  \rulename{ThreadAssoc}
}{
  T \conc (T' \conc T'')
}{
  (T \conc T') \conc T''
}

\crule{
  \rulename{ThreadGC}
}{
  \zerokt \conc T
}{
  T
}

\end{multicols}
\bigskip

Rule \rulename{ThreadInAgent} allows the use of the congruence rules
for threads in the layer of agent states:

\pcrule{
  \rulename{ThreadInAgent}
}{
  T \equiv T'
}{
  a(h, C, H, T \conc T'', W)
}{
  a(h, C, H, T' \conc T'', W)
}


\subsection{The Reduction Rules}
\label{sec:vm:rrules}

Each \mob instruction requires at least one machine transition to be
processed. The rules are written using the usual forms in the
definition of operational semantics.
The rules are of two forms: the first are denoted by 

\begin{step}
 \calA \reduces \calA'
\end{step}

\noindent
and operate simply over a pool of
threads. They are used whenever the operation to be performed does
not modify the name resolver $\ns$. 
The second, denoted by

\begin{step}
 \calA, \ns \reduces \calA', \ns'
\end{step}

\noindent
include the resolver and are used whenever the operation requests data
from the resolver or modifies its contents in some way.
To widen the scope of the first form of reductions to whole network we define rule
rule \rulename{AgentRed}. This allows us to define rules focused on one agent
alone, whenever the remainder of the network is not affected.

\pcrrule{
   \rulename{AgentRed}
}{
 \quad  \incnrule{\calA}{\calA''}
}{
  \calA \conc \calA', \ns
}{
 \calA'' \conc \calA', \ns
}

Rule \rulename{Cong} allows reduction to occur under structural
congruence:

\pcrrule{
   \rulename{Cong}
}{
   \calA \equiv \calA' \quad 
   \incnrule{\calA', \ns}{\calA'', \ns''} \quad
   \calA'' \equiv \calA''' \quad 
}{
  \calA', \ns 
} {
  \calA''', \ns''
}

Next we provide the rules for the language constructs.

\subsubsection*{Creation of Objects and Agents}

The instantiation of a regular class creates a new object. It reserves
a block of heap space for a closure representing the object. The
closure holds the values of the attributes, a special attribute
\selfk, that is a reference to the object itself, and keeps a link for the code
of the class.

\pnrule{
  \rulename{NewObject} 
}{
  \evalseq(H, t, B, \tilde v) = \tilde u \quad 
  C(X) = (\falsek, \tilde x, \_, \_, \epsilon) \quad
  r' \in \HeapRef\ \fresh
}{
  a(h, C, H, (t, (B, x = \newk\ X (\tilde v)\  P) :: Q, r) \conc T, W)
}{
  a(h, C, H+\{r':(\nullr, (\falsek, \{\selfk : r', \tilde x : \tilde u\},
  X))\}, (t, (B + \{x : r'\}, P) :: Q, r) \conc T, W)
}

Agents in \mob{} are similar to objects, but they have an execution
unit associated to them. A new agent is placed in the network's pool
of agents. In the beginning, its location is the same as the agent
that created it.  It is initiated with a heap containing its closure
at $\mathmob{r'}$ (as in the \rulename{NewObject} rule).  A new thread
is created to execute the code of the agent's \maink\ method with the
agent's environment $\mathmob{B'}$, given by the attributes and
\selfk.  \maink is a required method that defines the agent's initial
behavior, an approach common to many programming languages.  The
parent agent keeps a binding to the reference $\mathmob{r'@b}$ of the
created agent in $\mathmob{x}$.

The code repository for the new agent is
composed of the code required
by the values given as argument to the constructor ($\mathmob{C'}$),
plus all the code required by the agent definition ($\{\mathmob{X : C(X)}\}$).

\pppcnnrule{
  \rulename{NewAgent}
}{
  \evalseq(H, t, B, \tilde v) = \tilde u \quad 
  \copyseq_{ab}(C, H, \tilde u) = (C', H', \tilde u') \quad
  C(X) = (\truek, \tilde x, M, \_, S_1\ \cdots\ S_k)  
}{
  \codeOf(C(X), \maink) = (\epsilon, P') \quad
  B' = \{\selfk : r', \tilde x : \tilde u'\} \quad
  b \in \AgentName\ \text{and}\ r'@b \in \HeapRef\ \fresh
}{
  SNS(S_1) = (\alpha_1 , K_1)\ \cdots\
  SNS(S_k) = (\alpha_k , K_k) \quad  K_1 = \{ r@_i \conc i \in \{1,
  \dots, n\}\}\ \cdots\ K_k = \{ r@_i \conc i \in \{1, \dots, m\}\}
}{
  a(h, C, H, (t, (B, x = \newk\ X(\tilde v)\ P) :: Q, r) \conc T, W) \conc
  \calA, (ANS, SNS)
}{
  a(h, C, H, (t, (B+\{x:r'@b\}, P) :: Q, r) \conc T, W)\ \conc 
}{
  b(h, C' + \{ X : C(X)\}, H' + \{ r': (\nullr, (\truek, B', X))\}, \launchT(B', P', \nullr), \emptyset) \conc  \calA, 
}{
 (ANS + \{r'@b : h\}, SNS + \{S_1 : (\alpha_1, K_1 + \{r'@b\}),
 \dots, S_k: (\alpha_k, K_k + \{r'@b\})\}
}

\noindent
where $\mathmob{k}$ denotes the number of services implemented by the
agent, and $\mathmob{n}$ and $\mathmob{m}$ denote, respectively, the
number of implementations of services $\mathmob{S_1}$ and
$\mathmob{S_k}$ in the network.

Note that both maps of the resolver are updated.  The reference
$\mathmob{r'@b}$ holding the agent's closure will be the key in the
$\mathmob{ANS}$ map to locate the agent's current host $\mathmob{(r'@b
  : h)}$. Every entry of the $\mathmob{SNS}$ map corresponding to each
of the agent's implemented services given as $\mathmob K_1, \dots,
\mathmob K_n$, will be updated with $\mathmob{r'@b}$, e. g.,
$\mathmob{(S_1 : (\alpha_1, K_1 + \{r'@b\}))}$ for the implemented
service $\mathmob{S_1}$.

\subsubsection*{Multi-threaded Agents}

The \forkk{} instruction allows the explicit creation of a new thread
by the programmer. The new thread inherits the environment of its
creator and a handle is returned to the caller. This handle is
associated to a newly created heap reference, that contains a $\nullr$
value and is used for inter-thread synchronization.  The
synchronization is achieved by granting the thread exclusive access to
itself (see \rulename{Join} rules).

\pnrule{
  \rulename{Fork}
}{
  t' \in \ThreadRef\ \fresh
}{
  a(h, C, H, (t, (B, x = \forkk\ \{ P' \}\ ; P) :: Q, r) \conc T, W)
}{
  a(h, C, H + \{t' : (t', \nullr) \}, (t, (B+\{x:t'\}, P) :: Q, r) \conc
  (t', (B, P'), \nullr) \conc T, W) 
}

A thread can suspend waiting for the completion of another thread
using the instruction \joink{}. The instruction uses the thread's
handle returned by a previous \forkk{} statement. While it is running,
a given thread has a reference in the heap associated to it
($\mathmob{t'}$). In this scenario, any other thread that tries to
perform the \joink operation will suspend on $\mathmob{t'}$.

\prule{
   \rulename{JoinSuspend}
}{
  \evalone(H, t, B, x) = t' \quad 
    H(t') = (t', \nullr) \quad t' \not = t 
}{
  a(h, C, H, (t, (B, \joink(x)\ ; P) :: Q, r) \conc T, W) 
}{
  a(h, C, H, T, W + \{t' : (t, (B, P) :: Q, r)\})
}

If the thread on which the synchronization is performed is no longer
running, the reference associated to it is no longer locked. In this
case, the operation succeeds and the execution continues.

\prule{
   \rulename{Join}
}{
  \evalone(H, t, B, x) = t' \quad  (t = t'\ \vee\ H(t') = (\nullr, \nullr))
}{
  a(h, C, H, (t, (B, \joink(x)\ ; P) :: Q, r) \conc T, W) 
}{
  a(h, C, H, (t, (B, P) :: Q, r) \conc T, W) 
}

If the code currently under execution terminates and the stack has no
more elements, the thread has run out of code to execute and
terminates, as in rule \rulename{End}.  To give a more expressive
writing of the reduction rules involving synchronization, we define the
\notify macro. The macro represents a thread that wakes up all the
threads suspended on  reference $\mathmob{r}$.

\begin{center}
\macrorule{
}{
  \notify(r)
}{
  (\nullr, \epsilon, r)
}
\end{center}

When a thread terminates its execution  it uses
this macro to wake up all the threads suspended on it.  


\cnrule{
   \rulename{End}
}{
  a(h, C, H, (t, (B, \epsilon), \nullr) \conc T, W)
}{
  a(h, C, H, \notify(t) \conc T, W)
}

The \rulename{NotifyThread} rule wakes up every thread
suspended on the given reference. $\mathmob{W(r)}$ is the set of threads
suspended on the reference $\mathmob{r}$.

\cnrule{
   \rulename{NotifyThread}
}{
  a(h, C, H, \notify(r) \conc T, W)
}{
  a(h, C, H, T \conc \run(W(r)), W|_{\dom(W)- \{r\}})
}


Explicit synchronization is also supported in \mob by the \waitk and \notifyk instructionsm, that 
The first allows a thread to suspend on a reference, while the second is the language support to create to
a  \notify  thread, and thus wake every thread suspended on the given reference.

\prule{
   \rulename{Wait}
}{
  \evalone(H, t, B, x) = r' \quad  r' \not = t 
}{
  a(h, C, H, (t, (B, \waitk(x)\ ; P) :: Q, r) \conc T, W) 
}{
  a(h, C, H, T, W + \{r' : (t, (B, P) :: Q, r)\})
}

\prule{
   \rulename{Notify}
}{
  \evalone(H, t, B, x) = r'
}{
  a(h, C, H, (t, (B, \notifyk(x)\ ; P) :: Q, r) \conc T, W) 
}{
  a(h, C, H, \notify(r) \conc (t, (B, P) :: Q, r) \conc T, W)
}

\subsubsection*{Agent Movement and Discovery}

An agent may move to another host, changing the topology of the
distributed computation, rule \rulename{Go}. The original host will
proceed without the agent, and the later will resume its
execution concurrently with the agents at the target host. In order to
migrate an agent must be registered in the $\mathmob{ANS}$ map. 
The reference associated to the agent in $\Ans$ is discovered by following the
binding for the \selfk identifier. This can be done, since the \gok
instruction can only appear inside the body of an agent definition's
method.

\pnrule{
   \rulename{Go}
}{
   \evalone(H, t, B, v) = h' \quad 
  B(\selfk) = r'@a \quad
  r'@a \in \dom(ANS) \quad 
  h' \in \Host \quad
}{
   a(h, C, H, (t, (B,  \gok(v)\ ; P) :: Q, r) \conc T, W)\ |\ \calA, (ANS, SNS)
}{
   a(h', C, H, (t, (B,  P) :: Q, r) \conc T, W)\ |\ \calA, (ANS + \{r'@a : h'\}, SNS)
}

An agent may invoke a method in another agent only if it has a binding
for the target agent's closure.  Agent discovery in \mob is
service-oriented, meaning that agents are discovered for the services
they implement.  The instruction $\bindk$ consults the network
resolver and retrieves a heap reference, $\mathmob{r'@b}$, associated
with an agent that implements a service $\mathmob{S}$ and is presently
running in host $\mathmob{h}$. Note that an agent cannot obtain a
binding for itself.

\ppnrule{
   \rulename{Bind}
}{
   \evalone(H, t, B, v) = h' \quad
  \Sns(S) = (\alpha, \{r@a_1, \dots, r@a_n\}) 
}{
  \exists r'@b \in \{r@a_1, \dots, r@a_n\}\ :\ ANS(r'@b) = h'\ \wedge\ b \not = a
}{
   a(h, C, H, (t, (B,  x = \bindk(S\ v)\ ; P) :: Q, r) \conc T, W)\ |\ \calA, (\Ans, \Sns)
}{
   a(h, C, H, (t, (B + \{x : r'@b\},  P) :: Q, r) \conc T, W)\ |\ \calA, (\Ans, \Sns)
}

However, sometimes the host where the agent is running is irrelevant
and is not taken is consideration when the reference is picked. In
both these rules, the criteria used in choosing an agent is left to
the implementation.

\pnrule{
   \rulename{BindAny}
}{
  SNS(S) = (\alpha, \{r@a_1, \dots, r@a_n\}) \quad 
  \exists r'@b \in \{r@a_1, \dots, r@a_n\}\ :\ b \not = a
}{
   a(h, C, H, (t, (B,  x = \bindk(S)\ ; P) :: Q, r) \conc T, W)\ |\ \calA, (ANS, SNS)
}{
   a(h, C, H, (t, (B + \{x : r'@b\},  P) :: Q, r) \conc T, W)\ |\ \calA, (ANS, SNS)
}

\subsubsection*{Current Host}

The next rule returns the host where the agent is running.

\cnrule{
   \rulename{Host}
}{
   a(h, C, H, (t, (B,  x = \hostk()\ ; P) :: Q, r) \conc T, W)
}{
   a(h, C, H, (t, (B + \{x : h\},  P) :: Q, r) \conc T, W)
}

\subsubsection*{Local Method Invocation}

Method invocation in objects can only be done within an agent. All
objects are encapsulated within agents and thus, invoking a method in
an object located in the address space of some other agent is not
possible, unless the target agent's interface provides some means to
access the object.  Methods of the agent itself can of course be
invoked both from within the agent, and from other remote agents.
Rule \rulename{LocalInvoke} applies to the scenario of local
invocations, both in objects and in agents. We guarantee this
restriction by qualifying the reference with its location.

The method invocation simulates a call stack by suspending the current
thread, and by creating a new one, bound to the same heap reference
($\mathmob{t}$), to execute the body of the method.  The environment
of the new thread is obtained from the target object's environment
modified with the values assigned to the method's parameters.  The
result location is a fresh heap reference $\mathmob{r''}$, locked by the current thread
and holding no value $\mathmob{(r'' : (t, \nullr))}$.
The current thread is then suspended on that reference $\mathmob{r''}$, waiting for
the result. Its environment is modified by the binding of variable $\mathmob{x}$
to $\mathmob{r''}$, so that $\mathmob{x}$ holds the returned value once the current thread
resumes its execution.
By associating the new thread to the same heap reference as the one that invokes the method,
the former gains access to all the resources locked by the
later.

\uppnrule{
   \rulename{LocalInvoke}
}{
  \evalseq(H, t, B, \tilde v) = \tilde u \quad   
  B(o)= r'@a \quad
  \access(H, r'@a, t) = \truek \quad
  H(r'@a) = (\_, (\_, B', X))
}{
  \codeOf(C(X), m) = (\tilde{x}, P') \quad
  r'' \in \HeapRef\ \fresh
}{
  a(h, C, H, (t, (B, x = o.m(\tilde v)\ ; P) :: Q, r) \conc T, W)
}{
  a(h, C, H+\{r'' : (t, \nullr)\}, (t, (B' + \{\tilde x : \tilde
  u\}, P'), r''), W + \{ r'' : (t,  (B + \{x : r''\}, P) :: Q, r) \})
}

Note that elements of the heap of the form $\mathmob{r'' : (t,
  \nullr)}$ (with $\mathmob r \not \in \ThreadRef$ and $\mathmob t
\not = \nullr$) denote uniquely references waiting for results. This
will be important when encoding \mob to the target process calculus.

Rule \rulename{LocalInvokeLocked} states that if the object on which
the method its to be invoked is locked by another thread, the current
thread suspends on that object.

\uprule{
   \rulename{LocalInvokeLocked}
}{
  B(o)= r'@a \quad 
  \access(H, r'@a, t) = \falsek \quad
}{
  a(h, C, H, (t, (B, x = o.m(\tilde v)\ ; P) :: Q, r) \conc T, W)
}{
  a(h, C, H, T, W + \{r' : (t, (B, x = o.m(\tilde v)\ ; P) :: Q, r)\})
}

The \returnk instruction terminates the execution of the current
thread, places the result in the dedicated heap reference
$\mathmob{r}$, releasing its lock, and spawns a \notify thread to wake
up the thread waiting for the result.
Thus, the image of $\mathmob{r}$ in the heap will now hold the returned
value $\mathmob{(\nullr, u)}$.
The \notify thread will cause any thread in $\mathmob{W}$ waiting on
$\mathmob{r}$ to resume, simulating the call stack.

\uprule{
   \rulename{LocalReturn}
}{
  \evalone(H, t, B, v) = u 
}{
  a(h, C, H, (t, (B, \returnk(v)\ ; P) :: Q, r) \conc T, W) 
}{
  a(h, C, H + \{r : (\nullr, u)\}, \notify(r) \conc T, W) 
}

\subsubsection*{Remote Method Invocation in an Agent}

As in local method invocations, remote method invocations always
launch a new thread ($\mathmob{t'}$) in the target agent to execute
the corresponding code\footnote{From a practical point of view, the
  maximum number of threads allowed for one agent is
  implementation-dependent.  Note that remote invocations are not
  anonymous, the invoking agent may be identified, since the agent's
  name qualifies the reference $\mathmob{r''@a}$.  This means that
  precautions to avoid abusive use from other agents may be achieved
  by adding a set of new preconditions to the rule.  This, can be used
  for instance to avoid denial-of-service attacks.}.  The difference
lies in the fact that the result slot of the thread,
$\mathmob{r''@a}$, is now a heap reference from the heap of the
calling agent. Moreover, this thread in the case of remote
invocations, does not execute the body of the method, but rather
triggers a local invocation. This allows for the application of the
\rulename{LocalInvoke} reduction rule to execute the method locally at the remote agent.

The values assigned to the method's parameters are passed by value,
except agents that are passed by reference\footnote{Passing them by
  value would constitute a new form of migration.}. A copy of the
arguments must be sent to the target agent and since this may include
objects, a closure with the values and the classes they use must be
constructed, using function \copyseq.  The local invocation performed
at the target agent has arguments $\mathmob{\tilde x}$, bound to the
clones of the original values assigned to the arguments in the calling
thread.

\ppcnnrule{
   \rulename{RemoteInvoke}
}{
  \evalseq(H, t, B, \tilde v) = \tilde u \quad   
  B(o)= \at{r'}{b} \quad
  H'(\at{r'}{b}) = (\_, (\truek, B', X)) 
}{
  \copyseq_{ab}(C, H, \tilde u) = (C'', H'', \tilde u') \quad
   r''@a \in\ \HeapRef\  \fresh
}{
  (a(h, C, H, (t, (B,  x = o.m(\tilde v)\ ; P) :: Q, r) \conc\ T, W) \conc\ b(h', C', H', T', W')) \conc \calA, \ns
}{
  (a(h, C, H + \{r'' : (t, \nullr)\}, T, W + \{r'' : (t, (B + \{x :
  r''\}, P) :: Q, r) \}) \conc\ 
}{
  b(h', C' + C'', H' + H'', \launchT(\{\selfk: r', \tilde x : \tilde u'\}, x = \selfk.m(\tilde x) ; \returnk\ x, r''@a) \conc\  T', W')) \conc \calA, \ns
}



The return value from a remote method invocation must be placed in a
reference in the heap of the calling agent. This value may include
objects, and thus a closure with the value and the classes it uses
must be constructed.  Finally, a \notify thread is placed in the pool of
threads of the calling agent, that will trigger the \rulename{Notify}
rule and awake the  thread that performed the invocation.

\pnrule{
  \rulename{RemoteReturn}
}{
   \evalone(H, t, B, v) = u \quad  
  \copyone_{ab}(C, H, u) =  (C'',H'', u') 
}{
  (a(h, C, H, (t, (B, \returnk(v)\ ; P) :: Q, r@b) \conc T, W) \conc\
  b(h', C', H', T', W')) \conc \calA, \ns
}{
  (a(h, C, H, T, W) \conc
  b(h', C'+C'', H'+H''+\{r : (\nullr, u')\}, \notify(r) \conc T', W')) \conc \calA, \ns
}

\subsubsection*{Exclusive Access}

Values in the heap may be shared by several threads, therefore it is
necessary to supply a mechanism to ensure that a thread may gain
exclusive access to a given value.  The \lockk instruction gives
exclusive access to a reference to the current thread.  The operation
is only allowed if no other thread has exclusive access over the
reference. Note that although the inspected agent is the only one
under the rule's scope, we qualify $\mathmob{r'}$ with its location.
This is to point out that exclusive access operations can only be
performed on references owned by the agent.

\prule{
  \rulename{Lock}
}{
   \evalone(H, t, B, x) = r'@a \quad 
  \locked(H, r'@a, t) = (H', \truek)
}{
  a(h, C, H, (t, (B, \lockk(x)\ ; P) :: Q, r)\ |\  T, W)
}{
  a(h, C, H', (t, (B, P) :: Q, r)\ |\  T, W)
} 

A thread that tries to obtain the lock of a locked reference suspends on the
reference.

\uprule{
   \rulename{LockFailed}
}{
   \evalone(H, t, B, x) = r'@a \quad 
  \locked(H, r'@a, t) = (H, \falsek)
}{
  a(h, C, H, (t, (B, \lockk(x)\ ; P) :: Q, r)\  |\ T, W)
}{
  a(h, C, H, T, W +  \{r'@a : (t, (B, \lockk(x)\ ; P) :: Q, r)\})
}

Instruction \unlockk returns the public access to a given heap reference and
notifies every thread suspended on it, so that they may
resume their execution.

\uprule{
   \rulename{Unlock}
}{
     \evalone(H, t, B, x) = r'@a \quad 
  \unlocked(H, r'@a, t) = (H', \truek)
}{
  a(h, C, H, (t, (B,  \unlockk(x)\ ; P) :: Q, r)\ |\ T, W)
}{
  a(h, C, H', \notify(r'@a) \conc (t, (B, P) :: Q, r) \conc  T, W)
}

If a thread tries to
free an object without having exclusive access to it, the operation is
ignored.

\prule{
  \rulename{UnlockIgnore}
}{
   \evalone(H, t, B, x) = r'@a \quad 
  \unlocked(H, r'@a, t) = (H, \falsek)
}{
  a(h, C, H, (t, (B, \unlockk(x)\ ; P) :: Q, r) \conc  T, W)
}{
  a(h, C, H, (t, (B, P) :: Q, r)\ |\  T, W)
}

\subsubsection*{Control Flow}

The machine defines a basic set of instructions dedicated to control
the flow of execution (\ifk{}, \whilek{}, and \breakk).
The \ifk instruction requires two reduction rules, selecting the
branch according to the boolean value resulting of the evaluation of
value $\mathmob{v}$. Each of them executes the code of the selected branch
followed by the instruction's continuation ($\mathmob{P}$).

\prule{
  \rulename{IfTrue}
}{
  \evalone(H, t, B, v) = \truek
}{
  a(h, C, H, (t, (B, \ifk\ (v)\ \{ P' \}\ \elsek\ \{ P'' \}\ ; P) :: Q, r) \conc T, W)
}{
  a(h, C, H, (t, (B, P' ; P) :: Q, r) \conc T, W)
}

\prule{
   \rulename{IfFalse}
}{
  \evalone(H, r, B, v) = \falsek
}{
  a(h, C, H, (t, (B, \ifk\ (v)\ \{ P' \}\ \elsek\ \{ P'' \}\ ; P) :: Q, r) \conc T, W)
}{
  a(h, C, H, (t, (B, P'' ; P) :: Q, r) \conc T, W)
}

The \whilek instruction requires three rules. Rule \rulename{PushCont}
simply pushes the continuation of the instruction to the stack.  This
is required to allow the use of the \breakk instruction to branch out
of the loop (see rule \rulename{Break}). 

\ucnrule{
    \rulename{PushCont}
}{
  a(h, C, H, (t, (B, \whilek(v)\ \{ P' \} ; P) :: Q, r) \conc T, W)
}{
  a(h, C, H, (t, (B, P' ; \whilek(v)\ \{ P' \}) :: (B, P) ::  Q, r)  \conc\ T, W)
}

Rule \rulename{WhileTrue} executes the body of the \whilek instruction
composed with the instruction again, performing the loop.  The process
eventually stops when the value $\mathmob{v}$ evaluates to \falsek.
The execution then continues with the continuation popped from the
stack.

\uprule{
    \rulename{WhileTrue}
}{
  \evalone(H, t, B, v) = \truek
}{
  a(h, C, H, (t, (B, \whilek(v)\ \{ P \}) :: Q, r) \conc T, W)
}{
  a(h, C, H, (t, (B, P\ ; \whilek(v)\ \{ P \}) ::  Q, r)  \conc\ T, W)
}

Rule \rulename{WhileFalse} emulates the end of the loop resorting to
the \breakk instruction.

\prule{
   \rulename{WhileFalse}  
}{
  \evalone(H, t, B, v) = \falsek
}{
  a(h, C, H, (t, (B, \whilek\ (v)\ \{ P \}) ::  Q, r) \conc T, W)
}{
  a(h, C, H, (t,  (B, \breakk) :: Q, r) \conc T, W)
}

The \breakk instruction branches out of the loop. It pops the current
code-block from the stack and begins the execution of the continuation
(the new top of the stack). The environment of the continuation is
updated with the modifications performed during the execution of the
loop.


\ucnrule{
    \rulename{Break} 
}{
  a(h, C, H, (t, (B, \breakk\ ; P) :: (B', P') :: Q, r) \conc T, W)
}{
  a(h, C, H, (t, ((B' + B)|_{\dom(B')}, P') :: Q, r) \conc  T, W)
}






\subsubsection*{Execute External Services}

The \execk instruction allows the interaction with external services.
This interaction is defined by an interface of seven possible actions.

\begin{itemize}
\item \texttt{init}: opens a session with a service, and returns a session identifier;
\item \texttt{read}: reads a given number of bytes from a session;
\item \texttt{readLine}: reads a line from a session;
\item \texttt{write}: writes the given data to a session;
\item \texttt{action}: posts an action to be performed by the service associated to the session;
\item \texttt{isAlive}: checks if the session is still active;
\item \texttt{close}: closes the session.
\end{itemize}

The syntax of the \execk\ instruction requires the existence of three
arguments, of which the first is the string that determines which
action is to be performed.
The second argument is an integer value that, when the action is
\texttt{init}, corresponds to the identifier of the service to be
requested, and otherwise corresponds to the session identifier.
The third argument is a string used to pass values to the action.
The operation is performed by an internal built-in function
\textsf{exec} that executes synchronously.  Asynchronous calls can be
performed by encapsulating the \execk instruction in a new thread.

\uprule{
    \rulename{Exec}
}{
    \evalseq(H, t, B, v_1\ v_2\ v_3) = u_1\ u_2\ u_3 \quad
    \textsf{exec}(u_1\ u_2\ u_3) = u
}{
  a(h, C,H,(t, (B, x = \execk(v_1\ v_2\ v_3)\ ; P) :: Q, r) \conc T,W)
}{
  a(h, C, H,(t , (B + \{x : u\}, P) :: Q, r) \conc T,W)
}

The protocol to interact with an external service is initiated by the \texttt{init} action, that
receives as argument an integer value that identifies the service, and returns the session identifier. 
Once the session is opened, a series of \texttt{read},
\texttt{readLine}, \texttt{write}, \texttt{action}, and
\texttt{isAlive} actions may be performed.
To terminate the session, the \texttt{close} action must be used.
Below is an example of a session with a FTP server. We assume  that the
FTP service identifier is 4.

\begin{step}
\begin{array}{l}
\mathmob{x = \execk(\texttt{"init"}\ 4\ \texttt{"ftp.adomain"});}\\
\mathmob{x' = \execk(\texttt{"action"}\ x\ \texttt{"GET afile"});}\\
\mathmob{x' = \execk(\texttt{"read"}\ x\ \texttt{"4096"});}\\
\mathmob{y = x'\ != \texttt{""};}\\
\mathmob{\whilek\ (y)\ \{}\\
\quad \mathmob{x' = \execk(\texttt{"read"}\ x\ \texttt{"4096"});}\\
\quad \mathmob{y = x'\ != \texttt{""};}\\
\mathmob{\}}\\
\mathmob{x' = \execk(\texttt{"close"}\ x\ \texttt{""})}
\end{array}
\end{step}

The example begins by opening a FTP session with a server located at
\texttt{ftp.adomain}. The correspondent session identifier is placed
on $\mathmob{x}$. Next, it posts the \texttt{GET afile} action to fetch file
\texttt{afile}, and reads its contents in chunks of 4096 bytes.
Once the file is read, it closes the session.

\subsubsection*{Assignment of Expressions}

The result of the computation of an expression may be assigned to a variable. The assignment involves adding
a new entry in the environment ($\mathmob{B}$) of the thread.

\prule{
    \rulename{Assignment}
}{
    \evalone(H, t, B, e) = u
}{
  a(h, C,H,(t, (B, x = e\ ; P) :: Q, r) \conc T,W)
}{
  a(h, C,H,(t , (B + \{x : u\}, P) :: Q, r) \conc T,W)
}



\subsubsection*{Handling Attributes}

Assigning a value to an attribute of an object involves modifying the
object's closure. Thus, an inspection to the status of both the object
and the attribute is required. If the access to both is granted to the
current thread, the binding of the given attribute in the object's
closure is modified.

\uppnrule{
    \rulename{AttrAssignment}
}{
  \evalone(H, t, B, v) = u \quad
  B(\selfk) = r' \quad
  H(r') = (t', (bool, B', X)) \quad \access(H, r', t) =  \truek
}{
  (\evalone(H, t, B', x) = r'' \wedge \access(H, r'', t) =  \truek) \vee \evalone(H, t, B', x) = c
}{
  a(h, C, H,(t, (B, \selfk.x = v\ ; P) :: Q, r) \conc T,W)
}{
  a(h, C, H + \{r' : (t', (bool, B' + \{x : u\}, X))\},(t ,(B, P) :: Q, r) \conc T,W)
}

If not, the current thread may suspend on the reference holding the
object, or on the one holding the actual attribute. Rule
\rulename{AttrAssignmentLocked} covers the first case, where the
thread cannot access the object.

\uprule{
    \rulename{AttrAssignmentLocked}
}{
  B(\selfk) = r' \quad
  \access(H, r', t) =  \falsek
}{
  a(h, C, H,(t, (B, \selfk.x = v\ ; P) :: Q, r) \conc T,W)
}{
  a(h, C, H, T, W + \{r' : (t, (B, \selfk.x = v\ ; P) :: Q, r)\})
}

Rule \rulename{AttrAssignmentLockedInAttr} covers the second case,
where the thread has access to the object, but not to the attribute.

\upnrule{
    \rulename{AttrAssignmentLockedInAttr}
}{
  B(\selfk) = r' \quad
  H(r') = (t', (bool, B', X)) \quad
  \access(H, r', t) =  \truek
}{
  \evalone(H, t, B', x) = r'' \quad
  \access(H, r'', t) =  \falsek
}{
  a(h, C, H,(t, (B, \selfk.x = v\ ; P) :: Q, r) \conc T,W)
}{
  a(h, C, H, T, W + \{r'' : (t, (B, \selfk.x = v\ ; P) :: Q, r)\})
}

There is no access restriction on the reading of attributes. 
The rule simply retrieves the value of the attribute and binds the given variable to it.
To ensure that correctness of the information to be read, the
programmer must protect the access with a lock to the object.


\prule{
    \rulename{ReadAttr}
}{
  B(o) = r' \quad H(r') = (\_, (\_, B', X)) \quad B'(y) = u
}{
  a(h, C,H,(t, (B, x = \selfk.y\ ; P) :: Q, r) \conc T,W)
}{
  a(h, C,H,(t , (B + \{x : u\}, P) :: Q, r) \conc T,W)
}



\subsubsection*{Terminate an Agent}

Finally, the \rulename{Exit} rule terminates the execution of an
agent.  This is required because agents are daemons and their
execution must be explicitly terminated by the \exitk\ instruction.
All the references to the agent in the network must be removed.

\uppnrule{
    \rulename{Exit}
}{
  B(\selfk) = r \quad
  H(r) = (\_, (\_, \_, X)) \quad
  C(X) = (\_, \_, \_, \_, S_1\ \cdots\ S_n) \quad
}{
  SNS(S_1) = (\alpha_1 , \{r@a_{1_1}, \dots, r@a, \dots, r@a_{1_m}\})\ \dots\  SNS(S_n) = (\alpha_n , \{r@a_{n_1}, \dots, r@a, \dots, r@a_{n_k}\})
}{
  a(h, C, H,(t, (B, \exitk\ ; P) :: Q, r) \conc T,W) \conc \calA, (ANS, SNS)
}{
  \calA, (\Ans|_{\dom(\Ans) - \{a\}}, 
SNS + \{S_1 : (\alpha_1, \{r@a_{1_1}, \dots, r@a_{1_m}\}),  \dots, S_n : (\alpha_n, \{r@a_{n_1}, \dots, r@a_{n_k}\})\})
}


\section{The Type System}
\label{sec:lang:type-system}

In this section we present a type inference system for \mob that is
very much inspired in the type-system developed by Vasco Vasconcelos
for the \ptyco calculus \cite{phd:vasco:94}.
Types are ranged over by $\alpha$, and are distinguished between types
for primitive constants, ranged over by $\rho$, types for classes and agents,
types for objects, instances of agents and services, and a denumerable set of variables for types,
ranged over by $t$.
A type for a class (or agent) is defined by a tuple of two elements. The first
holds the types for class (or agent) attributes, and the second, of the form $\beta$,
defines the type for the interface of the class (or agent).
$\beta$ types are  records of the form $ \mathmob{\{m_1 : (\tilde
  \alpha_1\ \mapsto\ \alpha_1),\ ...,\ m_n : (\tilde \alpha_n\
  \mapsto\ \alpha_n)\}}$, where $\mathmob{m_i}$ denotes the identifier
of a method; $\tilde \alpha_i$, the types of its parameters, and;
$\alpha_i$, its return type.

\begin{small}
\begin{xalignat*}{2}
\alpha \grmeq & \rho\ & \text{Type of a primitive constant} \\
       \grmor & (\tilde \alpha, \beta) & \text{Type of a class or agent} \\
       \grmor &  \beta & \text{Type of an object,  an agent instance,
         or a service} \\
       \grmor & t\ & \text{Type variable}\\
       \grmor & \mu t.\alpha\ & \text{Type relational tree}\\
\beta \grmeq  & \mathmob{\{m_1 : (\tilde \alpha_1\ \mapsto\ \alpha_1),\ ...,\
  m_n : (\tilde \alpha_n\ \mapsto\ \alpha_n)\}} & \text{Record type}\\
\rho \grmeq   & \intt\ |\  \stringt\ |\ \booleant\ |\
\threadt & \text{Primitive types}\\
\end{xalignat*}
\end{small}

As in \ptyco, types are interpreted as rational (regular infinite)
trees.  A type denoted by $\mu t.\alpha$ with ($\alpha \not = t$)
represents the rational solution for the equation $\alpha = t$. An
interpretation of recursive types as infinite trees induces an
equivalence relation on types: $\alpha \approx \alpha'$, if the tree
solution for $\alpha = t$ and $\alpha' = t$ is the same.






\subsubsection*{Expressions}

\emph{Typings for expressions} are type assertions of the form 
$\mathmob{e} : \alpha$, for an expression $\mathmob{e}$, and its type
$\alpha$.
Expressions are formed by variables, constants, and operations over
both of these.  The type assignment is built from the types of constants and
built-in operations and of types assigned to variables. The type of
constants or a of built-in operation is given by the \typeoff built-in
function.  The later are represented as an application of the types of
the arguments into the type of the operation.  For example
$\typeof{\falsek} = \booleant$, and $\typeof{>} = \intt\ \intt\
\mapsto \booleant$.
%
For $\mathmob{\tilde v = v_1\ ...\ v_n}$, a sequence of pairwise distinct
values, and $\tilde \alpha = \alpha_1\ ...\ \alpha_n$, a sequence
of types, we denote $\mathmob{v_1 : \alpha_1,\ ...,\ v_n : \alpha_n}$, a
sequence of type assignments as: $\mathmob{\tilde v : \tilde \alpha}$.

\begin{multicols}{2}

\begin{step}
\rulename{Const}\quad \Gamma \vdash \mathmob{c : \typeof{c}}
\end{step}

\begin{step}
\rulename{Group}\quad\Gamma \vdash  \mathmob{e : \alpha \vdash (e) : \alpha}
\end{step}

\begin{step}
\rulename{Var}\quad \Gamma \cdot  \mathmob{x : \alpha \vdash x : \alpha}
\end{step}

\begin{step}
\rulename{Null} \quad
\Gamma \vdash \nullk : t && t\  \fresh \wedge \\
&& t \not \approx \rho \in \{\intt, \booleant, \stringt\}
\end{step}

\begin{step}
\rulename{Seqv}\quad 
\frac{
    \Gamma\ \vdash\   \mathmob{v_1 : \alpha_1}\ \dots\ \Gamma\ \vdash\  \mathmob{v_n : \alpha_n}
}{
    \Gamma\ \vdash\   \mathmob{\tilde{v} : \tilde{\alpha}}
}
\end{step}

\begin{step}
  \rulename{UnOp}\quad \mathmob{
    \frac{
      \mathmob{\typeof{uop} = \rho_1 \reduces \rho_2 \quad \Gamma\ \vdash\
        e : \rho_1}
    }{
      \Gamma\ \vdash\ \mathmob{uop\ e : \rho_2} 
    }
}
\end{step}
\end{multicols}

\begin{step}
\rulename{BinOp}\quad \frac{
  \mathmob{\typeof{bop} = \rho_1 \rho_2 \reduces \rho_3 \quad
  \Gamma\ \vdash\ e_1 : \rho_1 \quad
  \Gamma\ \vdash\ e_2 : \rho_2}
}
{
  \mathmob{\Gamma\ \vdash\ e_1\ bop\ e_2 : \rho_3}
} 
\end{step}

\subsubsection*{Instructions}

\emph{Type assignments for agents and services} are, correspondingly,
type assertions of the form $\mathmob{X} : (\tilde \alpha, \beta)$ or
$\mathmob{S} : \beta$. Typings, denoted by $\Gamma$,  is a map defined as:

\begin{step}
  \Gamma : (\Class \cup \ServiceId \cup \Var) \mapsto \Type
\end{step}

\noindent
that contains the type assigments for classes, agents,
services, and variables.
For $\gamma$ ranging over the elements of $\dom(\Gamma)$, we have that
$\Gamma \backslash \tilde\gamma$ denotes the typing obtained by
$\Gamma$ with its domain reduced from the elements in $\tilde \gamma$,
and $\Gamma(\gamma)$ as the typing assigned to $\gamma$ if $\gamma \in
\dom(\Gamma)$.
%
We also denote as $\mathmob{x_{ret_i}}$ as
the built-in identifier that holds the return type of a method
$\mathmob{m_i}$.

\emph{Type assignments for definitions, sequences of instructions and methods} are
thus denoted, respectively, by $\Gamma \vdash \mathmob{\tilde D}$, $\Gamma \vdash \mathmob{P}$ and $\Gamma
\vdash \mathmob{M}$.  We begin by presenting the rules for service
definitions and service requirement. Services are not removed from the
set of typings until they are checked against the types defined for
them in the network.
Thus, when the local inference is done, the name resolver
($\mathmob{\ns}$) is contacted
to validate the local typings for the services.

\begin{step}
\rulename{Service} \quad
\frac{
     \Gamma \vdash \mathmob{\tilde D} \quad
  \Gamma \vdash \mathmob{P}
}
{
 \Gamma \vdash \mathmob{\servicek\ S\ \{ m_1\ \dots\
    m_n \}\ \tilde D\ P} 
} \quad (\Gamma(\mathmob{S}) \approx \mathmob{ \{m_1 : (\tilde \alpha_1 \mapsto \alpha_1),\
    \dots,\ m_n : (\tilde \alpha_n \mapsto \alpha_n)\}})
\end{step}

\begin{step}
\rulename{Requires} \quad
\frac{
     \Gamma \vdash \mathmob{\tilde D} \quad
     \Gamma \vdash \mathmob{P}
}
{
 \Gamma \vdash \requiresk\ \mathmob{\tilde S\ \tilde D\ P} 
} \quad      (\Gamma(\mathmob{S_1}) \approx  \beta_1\ \cdots\  
     \Gamma(\mathmob{S_n}) \approx \beta_n)
\end{step}

\begin{small}
\rulename{ServiceCheck} \quad
\begin{tabular}{c}
$    \{ \mathmob{S_1} : \beta_1, \dots,  \mathmob{S_n} : \beta_n\}  \vdash \mathmob{\tilde D\ P} \quad
    \ns = (\Ans, \Sns)$\\  $\Sns(\mathmob{S_1}) =  (\beta'_1, \_) \cdots\ 
    \Sns(\mathmob{S_n}) =  (\beta'_n, \_) \quad
     \beta_1 \approx \beta'_1 \cdots\ \beta_n \approx\ \beta'_n$\\
\hline
$ \mathmob{\emptyset\ \vdash\ \tilde D\ P}$
\end{tabular}
\end{small}

Regarding classes and agents we define rules to type collections of
methods, class and agent definitions. To allow mutual recursion
between class and agent definitions we define two rules for both. One
applied in the general case, and one other only applied when the
definition is the last in the sequence. The later closes the system
for definitions, removing them from the set of bindings by using a
\textsl{defs} function that, given an set $\Gamma$, returns a sequence
of all the elements from its domain that belong to $\Class$.

\begin{step}
\begin{array}{l}
\rulename{MethodCollection} \\
\begin{tabular}{c}
    $\Gamma \vdash   \mathmob{\tilde x_1 : \tilde \alpha_1} \quad
    \Gamma \vdash  \mathmob{x_{ret}}_1 : \alpha_1 \quad
    \Gamma \vdash \mathmob{P}_1\quad
    \cdots\quad
    \Gamma \vdash  \mathmob{\tilde x_n} : \tilde \alpha_n \quad
    \Gamma \vdash  \mathmob{x_{ret}}_n : \alpha_n \quad
    \Gamma \vdash \mathmob{P}_n
  $\\ 
  \hline
  $
     \Gamma\  \backslash\ \mathmob{
     \tilde x_1\ {x_{ret}}_1\ \cdots\ \tilde x_n\ {x_{ret}}_n\ \vdash\  m_1(\tilde x_1)\ \{ P_1 \}\ \cdots\
     m_n(\tilde x_n)\  \{  P_n \}\ :\ \{m_1 : \tilde \alpha_1\ \mapsto\ \alpha_1, \dots, m_n : \tilde
    \alpha_n\ \mapsto\ \alpha_n\}
  }$
\end{tabular}
\end{array}
\end{step}

\begin{step}
\begin{array}{l}
\rulename{Class} \quad
\begin{tabular}{c}
$\Gamma \cdot \mathmob{\selfk : \beta \cdot
    \selfk.x_1 : \alpha_1 \cdot\ \cdots\  \cdot \selfk.x_n : \alpha_n \vdash M : \beta}\ \quad
   \Gamma \vdash \mathmob{\tilde D} \quad
   \mathmob{\tilde D} \not = \epsilon \quad
   \Gamma \vdash \mathmob{P} 
  $\\
\hline
$
\Gamma \vdash\ \mathmob{\classk\ X(\tilde x)\ M\ \tilde D\ P
  }$
\end{tabular}
\end{array}
\end{step}

\begin{step}
\begin{array}{l}
\rulename{ClassIsLastDef} \quad
\begin{tabular}{c}
$\Gamma \cdot \mathmob{\selfk : \beta \cdot
    \selfk.x_1 : \alpha_1 \cdot\ \cdots\  \cdot \selfk.x_n : \alpha_n \vdash M : \beta}\ \quad
   \Gamma \vdash \mathmob{P} 
  $\\
\hline
$
\Gamma  \backslash \textsl{defs}(\Gamma) \vdash\ \mathmob{\classk\ X(\tilde x)\ M\ P
  }$
\end{tabular}
\end{array}
\end{step}

\begin{step}
\begin{array}{ll}
\rulename{Agent} \quad
\begin{tabular}{c}
  $\Gamma \cdot \mathmob{\selfk : \beta \cdot
    \selfk.x_1 : \alpha_1 \cdot\ \cdots\  \cdot \selfk.x_n : \alpha_n \vdash M : \beta}\ \quad
   \Gamma \vdash \mathmob{\tilde D} \quad
   \mathmob{\tilde D} \not = \epsilon \quad
   \Gamma \vdash \mathmob{P} 
  $\\
  \hline
  $  \Gamma\ \vdash\ \mathmob{
\agentk\ X(\tilde x)\ \implementsk\ \tilde S\
    \requiresk\ \tilde S'\ M\ \tilde D\ P  
  }$ \\
  $(
    \Gamma(X) \approx   \mathmob{(\tilde \alpha, \beta),
    \forall S \in \tilde S: \Gamma(S) \approx  \beta' \implies \forall \mathmob m \in \dom(\beta'): \mathmob m \in \dom(\beta) \wedge \beta'(m) \approx \beta(m)
  }
  )$
\end{tabular}
\end{array}
\end{step}

\begin{step}
\begin{array}{ll}
\rulename{AgentIsLastDef} \quad
\begin{tabular}{c}
  $\Gamma \cdot \mathmob{\selfk : \beta \cdot
    \selfk.x_1 : \alpha_1 \cdot\ \cdots\  \cdot \selfk.x_n : \alpha_n \vdash M : \beta}\ \quad
   \Gamma \vdash \mathmob{P} 
  $\\
  \hline
  $  \Gamma\ \backslash \textsl{defs}(\Gamma) \vdash\ \mathmob{
\agentk\ X(\tilde x)\ \implementsk\ \tilde S\
    \requiresk\ \tilde S'\ M\ P  
  }$ \\
  $(
  \mathmob{
\forall S \in \tilde S: \Gamma(S) \approx \beta' \implies \forall \mathmob m \in \dom(\beta'): \mathmob m \in \dom(\beta) \wedge \beta'(m) \approx \beta(m)
  }
  )$
\end{tabular}
\end{array}
\end{step}

We now define the rules to type sequences of \mob instructions ($\mathmob{P}$):

\begin{multicols}{2}

\begin{step}
\rulename{Fork} \quad
\frac{
  \Gamma \vdash \mathmob{x : \threadt} \quad
   \Gamma \vdash \mathmob{P}' \quad
   \Gamma \vdash \mathmob{P}
}{
  \Gamma\ \backslash\ \mathmob{x\ \vdash\ x = \forkk \{ P' \}; P}
}
\end{step}

\begin{step}
\rulename{Wait} \quad
\frac{
  \Gamma \vdash\ \mathmob{x : \beta}\quad
   \Gamma \vdash \mathmob{P}
}
{
   \Gamma \vdash\ \waitk\ \mathmob{(x); P}
}
\end{step}

\begin{step}
\rulename{Lock} \quad
\frac{
  \Gamma \vdash\ \mathmob{x : \beta}\quad
   \Gamma \vdash \mathmob{P}
}
{
   \Gamma \vdash\ \lockk\ \mathmob{(x); P}
}
\end{step}

\begin{step}
\rulename{Host} \quad
\frac{
    \Gamma \vdash   \mathmob{x : \stringt} \quad
   \Gamma \vdash \mathmob{P}
}{
 \Gamma\ \backslash\ \mathmob{x\ \vdash\ x = \hostk(); P}
}
\end{step}

\begin{step}
\rulename{Expr} \quad
\frac{
    \Gamma \vdash   \mathmob{x : \alpha} \quad
    \Gamma \vdash   \mathmob{e : \alpha} \quad
   \Gamma \vdash \mathmob{P}
}{
  \Gamma\ \backslash\ \mathmob{x\ \vdash\ x = e; P}
}
\end{step}

\begin{step}
\rulename{While} \quad
\frac{
    \Gamma \vdash\   \mathmob{v : \booleant} \quad 
     \Gamma \vdash\ \mathmob{P'} \quad
   \Gamma \vdash \mathmob{P} 
  }
{
 \Gamma \vdash\   \mathmob{\whilek\ (v)\ \{ P' \}; P}
}
\end{step}

\begin{step}
\rulename{Exit} \quad
\frac{
   \Gamma \vdash \mathmob{P}
}
{
   \Gamma \vdash\ \mathmob{\exitk; P}
}
\end{step}

\begin{step}
\rulename{Join} \quad
\frac{
  \Gamma \vdash\ \mathmob{x} : \threadt \quad
   \Gamma \vdash \mathmob{P}
}
{
   \Gamma \vdash\ \joink\ \mathmob{(x); P}
}
\end{step}

\begin{step}
\rulename{Notify} \quad
\frac{
  \Gamma \vdash\ \mathmob{x : \beta}\quad
   \Gamma \vdash \mathmob{P}
}
{
   \Gamma \vdash\ \notifyk\ \mathmob{(x); P}
}
\end{step}

\begin{step}
\rulename{Unlock} \quad
\frac{
  \Gamma \vdash\ \mathmob{x : \beta} \quad
   \Gamma \vdash \mathmob{P}
}
{
   \Gamma \vdash\ \unlockk\ \mathmob{(x); P}
}
\end{step}

\begin{step}
  \rulename{Go} \quad
\frac{
  \Gamma \vdash\ \mathmob{v} : \stringt \quad
   \Gamma \vdash \mathmob{P}
}
{
   \Gamma \vdash\ \mathmob{\gok\ (v); P}
}
\end{step}

\begin{step}
\rulename{If} \quad
\frac{
  \mathmob{\Gamma \vdash\ v : \booleant} \quad
   \Gamma \vdash\ \mathmob{P'} \quad
   \Gamma \vdash\ \mathmob{P''} \quad
   \Gamma \vdash \mathmob{P}
}{
   \Gamma \vdash\ \mathmob{\ifk\ (v)\ \{ P' \}\ \elsek\ \{ P'' \}; P}
}
\end{step}

\begin{step}
\rulename{Break} \quad
\frac{
   \Gamma \vdash \mathmob{P}
}
{
   \Gamma \vdash\ \mathmob{\breakk; P}
}
\end{step}

\begin{step}
\rulename{Return} \quad
\frac{
  \Gamma \vdash\ \mathmob{v} : \alpha \quad \Gamma \vdash\ \mathmob{x_{ret}} : \alpha \quad
   \Gamma \vdash \mathmob{P}
}
{
   \Gamma \vdash\ \returnk\ \mathmob{v; P}
}
\end{step}
\end{multicols}

\begin{step}
\rulename{Bind} \quad
\frac{
\Gamma \vdash   \mathmob{x : \beta} \quad
    \Gamma \vdash   \mathmob{v : \stringt}  \quad
   \Gamma \vdash \mathmob{P}
}{
 \Gamma\ \backslash\  \mathmob{x\ \vdash\ x = \bindk(S\ v); P}
} \quad (\Gamma  \mathmob{(S) \approx  \beta})
\end{step}

\begin{step}
\rulename{BindAny} \quad
\frac{
 \Gamma \vdash  \mathmob{x : \beta}  \quad
     \Gamma \vdash \mathmob{P}
}{
  \Gamma\ \backslash\  \mathmob{x\ \vdash\ x = \bindk(S); P}
} \quad (\Gamma  \mathmob{(S) \approx  \beta})
\end{step}


\begin{step}
\rulename{New} \quad
\frac{
\Gamma \vdash   \mathmob{x : \beta} \quad
     \Gamma \vdash \mathmob{P}
}{
 \Gamma\ \backslash\  \mathmob{x\ \vdash\ x = \newk\ X(\tilde v); P}
}\quad (
\Gamma(\mathmob X) \approx   (\tilde \alpha, \beta) \quad
\Gamma(\mathmob{\tilde v}) \approx \tilde \alpha
)
\end{step}

\begin{step}
\rulename{MethodInv} \quad
\frac{
   \Gamma \vdash \mathmob{P}
}
{
   \Gamma\ \backslash\ \mathmob{x\ \vdash\ x = o.m(\tilde v); P}
} \quad 
 (
   \Gamma(\mathmob{o}) \approx \{..., \mathmob{m} : (\tilde \alpha\ \mapsto\ \alpha), ...\} \quad
   \Gamma(\mathmob{\tilde v}) \approx \tilde \alpha \quad
   \Gamma(\mathmob{x}) \approx \alpha) 
\end{step}

\begin{multicols} 2
\begin{step}
\rulename{AttrRead} \quad
\frac{
    \Gamma \vdash  \mathmob{x : \alpha} \quad
     \Gamma \vdash \mathmob{P}
}{
 \Gamma\ \backslash\ \mathmob{x\ \vdash\ x = o.y; P}
} \quad     (\mathmob{\Gamma(o.y) \approx \alpha})
\end{step}

\begin{step}
\rulename{AttrWrite} \quad
\frac{
    \Gamma \vdash  \mathmob{x : \alpha} \quad
     \Gamma \vdash \mathmob{P}
}{
 \Gamma \vdash\ \mathmob{o.y = x; P}
} \quad     (\mathmob{\Gamma(o.y) \approx \alpha})
\end{step}
\end{multicols}

\begin{step}
\rulename{ExecInt} \quad
\frac{
\mathmob{ 
  \Gamma \vdash x : \intt \quad
  \Gamma \vdash v_1 : \stringt \quad
  \Gamma \vdash v_2 : \intt \quad
  \Gamma \vdash v_3 : \stringt \quad
  v_1 = "init"
} \quad
   \Gamma \vdash \mathmob{P}
}{
 \Gamma\ \backslash\ \mathmob{x\ \vdash\ x = \execk(v_1\ v_2\ v_3); P}
}
\end{step}

\begin{step}
\rulename{ExecString} \quad
\begin{tabular}{c}
  $\mathmob{
  \Gamma \vdash x : \stringt \quad
  \Gamma \vdash v_1 : \stringt \quad
  \Gamma \vdash v_2 : \intt \quad
  \Gamma \vdash v_3 : \stringt}$ \\
  $\mathmob{(v_1 = "read" \vee v_1 =  "readLine")} \quad
   \Gamma \vdash \mathmob{P}$\\
  \hline
  $  \Gamma\ \backslash\ \mathmob{x\ \vdash\ x = \execk(v_1\ v_2\ v_3); P}$
\end{tabular}
\end{step}

\begin{step}
\rulename{ExecBool} \quad
\begin{tabular}{c}
$\mathmob{ 
  \Gamma \vdash x : \booleant \quad
  \Gamma \vdash v_1 : \stringt \quad
  \Gamma \vdash v_2 : \intt \quad
  \Gamma \vdash v_3 : \stringt}$ \\
$\mathmob{(v_1 = "write" \vee v_1 = "isAlive" \vee v_1 = "action \vee
  v_1 = "close")}\quad
   \Gamma \vdash \mathmob{P}$\\
  \hline
  $  \Gamma\ \backslash\ \mathmob{x\ \vdash\ x = \execk(v_1\ v_2\ v_3); P}$
\end{tabular}
\end{step}\\

Next, we present two simple programming examples in \mob and execute
one of them in the \mobam.

\section{Programming in \mob}
\label{sec:lang:examples}
\lstset{numbers = left, numberblanklines = false} 

We exemplify the syntax with two small examples.
We assume that two classes were previously defined. These are
\texttt{Array} and \texttt{Map}, and implement the usual operations
with arrays and maps. \texttt{FILEEXEC} and \texttt{IO} are two
integer constants that we also assume that were previously defined.
Besides the \texttt{Array} and \texttt{Map} classes, these examples
resort only to the base core \mob constructs, hence their verbosity. 
%

The first example is that of a server
and a client for a clock synchronising service (\texttt{Time}). The
server in listing \ref{lst:lang:times} provides a service that
features a single method \texttt{getTime()} (lines 5 to 10).  Note
that the \maink method in line 4 may be empty since \mob agents run as
daemons and some external action is required to terminate their
execution.  The program, not the launched agent, terminates with the
\exitk instruction at line 13.

\lstset{language=mob}
\lstset{backgroundcolor=,rulecolor=,showstringspaces=false,
  basicstyle=\ttfamily, keywordstyle=\sffamily\textbf,
  commentstyle=\itshape\sffamily, frame=tb, texcl}
\begin{scriptsize}
\begin{lstlisting}[caption=A time server agent, label=lst:lang:times]
service Time { getTime }

agent TimeServer() provides Time {

  main { }

  getTime() {
    d = exec("init", FILEEXEC, "getTimeApplication");  // Open the session
    x = exec("readLine", d, "");                       // Read the output of the application
    status = exec("close", d, "");                     // Close the session
    return (x);
  }
}
x = new TimeServer();                                  // Create agent
exit;                                                  // Terminate program
\end{lstlisting}
\end{scriptsize}

The client (listing \ref{lst:lang:timec}) requires the
$\mathmob{Time}$ service in line 1 and, when run,  takes an array of
hosts and performs a cycle (lines 7 to 16) in which it moves to each
of them in line 9, setting their clock according with central
time from the \texttt{TimeServer} (lines 10 to 13).
Lines 19 to 22 construct the array to be passed as argument to the
instance of the agent created in line 23. The program terminates its
execution in line 24.

\begin{scriptsize}
\begin{lstlisting}[caption=A time client agent, label=lst:lang:timec]
agent TimeClient(hostList) requires Time {

  main() { 
    timeServer = bind(Time);                   // Discover service
    iter = hostList.iterator();                // Build condition for while
    hasNext = iter.hasNext();
    cond = hasNext == true; 
    while(cond) {                              // For all hosts migrate and execute setTimeApplication
      hostName = iter.next();                  // Next host
      go(hostName);                            // Go to the next host
      time = timeServer.getTime();             // Get time from server  
      command = "setTimeApplication " ^ time;  // Build command to execute
      d = exec("init", FILEEXEC, command);     // Open session to execute the application
      status = exec("close", d, "");           // Close session 
      hasNext = iter.hasNext();                // Build condition for while 
      cond = hasNext == true; 
    }
  }
}
hosts = new Array(null, 0);                    // Construct array
x = hosts.put("host1.net1");
x = hosts.put("host2.net2");
x = hosts.put("host3.net3");
x = new TimeClient(hosts);                     // Create agent
exit;                                          // Terminate program
\end{lstlisting}
\end{scriptsize}

Another, slightly more complex application is a minimal
\texttt{Messenger} service implemented in listing
\ref{lst:lang:messs}.
The \texttt{Messenger} service, defined in line 1, provides three
methods: a client may log in the system (\texttt{logIn}), log out from
the system (\texttt{logOut}) or
ask who is currently on-line (\texttt{getLogged}). Their implementation is done
respectively in lines, 4 to 7, 8 to 11, and 12 to 14. The instance of
the agent is created in line 17, with the map created in the line
before. The programs terminates its execution in line 18.

\begin{scriptsize}
\begin{lstlisting}[caption=A messenger server agent, label=lst:lang:messs]
service Messenger {logIn logOut getLogged }

agent MessengerServer(logged) provides Messenger {

  logIn(nickname, client) {            // Log in the system
    x = logged.add(nickname, client); 
    return (null);
  }

  logOut(nickname) {                   // Log out from the system
    x = logged.remove(nickname); 
    return (null);
  }

  getLogged() {                        // Ask who is on-line
    return (logged);
  }
}
logged = new Map(null, 0);             // Create initial empty map
x = new MessengerServer(logged);       // Create agent
exit;                                  // Terminate program
\end{lstlisting}
\end{scriptsize}

The particular messenger client in this example (listing
\ref{lst:lang:messc}) first binds to the service and logs in the
system (lines 7
to 9). Then, it initiates an input/output service (lines 10 and 11) and starts a loop (lines 12 to 35) in which the people
currently on-line are listed (lines 13 to 23) and waits for the
nickname in the input at line 24.  The input triggers the creation of
a new session between the client and the selected peer. The session is
handled by a dedicated thread, which allows for several simultaneous
conversations (lines 25 to 33).  During the session, any input from
the keyboard of the client is sent to the receptor (lines 29 to 33),
until the \textsf{"quit"} keyword is typed to end the session.

A client provides the \textsf{MessengerPeer} service, defined in
line 1, to provide the method that receives and prints remote
messages (lines 38 to 41). A client is terminated by some event that invokes the
\textsf{close} method (lines 43 to 47) that logs out from the system and halts the
agent's execution.

\begin{scriptsize}
\begin{lstlisting}[caption=A messenger client agent, label=lst:lang:messc]
service MessengerPeer { printMessage }

agent MessengerClient(nickname, server, peersMap, io) 
      provides MessengerPeer requires Messenger {

  main() {
    aux = bind(Messenger);                   // Discover service
    self.server = x;                         // Assign it to the server attribute
    x = server.logIn(nickname, self);        // Log in the server
    aux = exec("init", IO, "");              // Open standard input/output session
    self.io = aux;                           // Assign it to the io attribute
    while (true) {
      aux = server.getLogged();              // Obtain peers logged in the server
      self.peersMap = aux;                   // Assign them to the peersMap attribute
      iter = peersMap.iterator();            // Obtain an iterator over the map of peers
      hasNext = iter.hasNext();              // Build condition for while
      cond = hasNext == true; 
      while(cond) { 
        p = iter.next();                     // Next peer 
        x = exec("write", io, p);            // Write the names of the people logged on the server
        hasNext = iter.hasNext();            // Build condition for while
        cond = hasNext == true; 
      }
      chosen = exec("readLine", io, "");     // Read the name of the peer selected for conversation
      fork { 
        peer = peersMap.get(chosen);         // Read first message to be sent
        line = exec("readLine", io, "");
        cond = line  != "quit";
        while (cond) {                       // While in conversation send message and read next
          dummy = peer.printMessage(line); 
          line = exec("readLine", io, "");
          cond = line  != "quit";
        }
      } 
    }
  }

  printMessage(line) { 
    x = exec("write", io, line);             // Print message in the screen
    return (null);
  }  

  close() {
    x = server.logOut(nickname);             // Send logout message to the server
    status = exec("close", io, "");          // Close input/output session
    exit;                                    // Terminate agent 
  }
}
x = new MessengerClient("nick", null, null, null);     // Create agent
exit;                                                  // Terminate program
\end{lstlisting}
\end{scriptsize}

\section{Executing an Example}

We are now going to exemplify how a \mob script its executed by the
\mobam. Our case studies will be the $\textsf{TimeServer}$ and
$\textsf{TimeClient}$ examples. We will only provide a partial
execution of both.
We will focus on the creation of the agents and on their interaction,
skipping some of the steps that are not directly related with these
operations, and terminating our execution once the interaction is over.

Scripts are executed in the \mobam by encapsulating them in agents
that execute their code.  We assume that the \mob network target of
our example is denoted by: $\calA, \ns$.  In the sequel,  we \underline{underline} the components that
have been altered by the application of a rule of the \mobam.



We begin by launching the $\textsf{TimeServer}$ script from listing
\ref{lst:lang:times}. We create a fresh key ($\mathmob a$) for the
agent running the script.  The configuration of the network once we
launch the agent is:

\begin{greduce}
   \mathmob{a}(
     \mathmob{h, C, \launchT(\emptyset,  x = \newk\ TimeServer();\ \exitk, \nullr),\emptyset}
     ) \conc \calA, \ns & \reduces
\end{greduce}

where $\mathmob{C = \{TimeServer : (\truek, \epsilon, \{\maink :
(\epsilon, \epsilon), getTime : (\epsilon, P)\}, \emptyset, Time)\}}$ is the
result of the application of function \codeclo to the program. 
Unfolding the \launchT macro we have:

\begin{greduce}
   \mathmob{a}(
     \mathmob{h, C, \underline{\{t : (t, \nullr)\}}, \underline{(t, (\emptyset,  x = \newk\ TimeServer();\ \exitk), \nullr)},\emptyset}
     ) \conc \calA, \ns & \reduces
\end{greduce}

Applying rule \rulename{NewAgent} we create the new
$\mathmob{TimeServer}$ agent.  Since rule \rulename{NewAgent} is 
somewhat extensive we present it again.

\pppcnnrule{
  \rulename{NewAgent}
}{
  \evalseq(H, t, B, \tilde v) = \tilde u \quad 
  \copyseq_{ab}(C, H, \tilde u) = (C', H', \tilde u') \quad
  C(X) = (\truek, \tilde x, M, \_, S_1\ \dots\ S_k)  
}{
  \codeOf(C(X), \maink) = (\epsilon, P') \quad
  B' = \{\selfk : r', \tilde x : \tilde u'\} \quad
  b \in \AgentName\ \text{and}\ r'@b \in \HeapRef\ \fresh
}{
  SNS(S_1) = (\alpha_1 , K_1)\ \cdots\
  SNS(S_k) = (\alpha_k , K_k) \quad  K_1 = \{ r@_i \conc i \in \{1,
  \dots, n\}\}\ \cdots\ K_k = \{ r@_i \conc i \in \{1, \dots, m\}\}
}{
  a(h, C, H, (t, (B, x = \newk\ X(\tilde v)\ P) :: Q, r) \conc T, W) \conc
  \calA, (ANS, SNS)
}{
  a(h, C, H, (t, (B+\{x:r'@b\}, P) :: Q, r) \conc T, W)\ \conc 
}{
  b(h, C' + \{ X : C(X)\}, H' + \{ r': (\nullr, (\truek, B', X))\}, \launchT(B', P', \nullr), \emptyset) \conc  \calA, 
}{
 (ANS + \{r'@b : h\}, SNS + \{S_1 : (\alpha_1, K_1 + \{r'@b\}),
 \dots, S_k: (\alpha_k, K_k + \{r'@b\})\}
}

For the sake of readability we  assume that there is
no implementation of the $\mathmob{Time}$ service in this network, and
thus that $\mathmob{ \Sns(Time) = (\alpha_1 , \emptyset)}$.
We know that the code for \maink is $\epsilon$, and that
$\mathmob{TimeServer}$ has no attributes. Thus, we do not require the
cloning of the attributes. The code for the new agent $\mathmob{b}$ is
simply the code for the class itself, meaning
$\mathmob C$. Thus,
by applying rule \rulename{NewAgent} we obtain:

\begin{gather*}
\begin{small}
\begin{array}{rl}
   \mathmob{a(h, C, \{t : (t, \nullr)\}, (t, (\underline{\{x : r@b\}}, \exitk), \nullr), \emptyset)}
     \conc  &\\
\mathmob{\underline{b(h, C, \{ r: (\nullr, (\truek, \{\selfk : r\}, TimeServer))\},
     \launchT(\{\selfk : r\}, \epsilon, \nullr), \emptyset)}} \conc \calA, &\\
     \mathmob{ \underline{(\Ans + \{r@b : h\}, \Sns + \{Time : (\alpha_1, \{r@b\})\})}} &\reduces
\end{array}
\end{small}
\end{gather*}

We may now use the \rulename{Exit} rule to terminate the execution of
the script (in agent $\mathmob{a}$), and rule \rulename{AgentGC} to garbage collect
the resulting $\zeroka$ agent.

\begin{greduce}
\underline{\zeroka} \conc
\mathmob{b(h, C, \{ r : (\nullr, (\truek, \{\selfk : r\}, TimeServer))\},
     \launchT(\{\selfk : r\}, \epsilon, \nullr), \emptyset) \conc
     \calA}, &\\
     \mathmob{(\Ans + \{r@b : h\}, \Sns + \{Time : (\alpha_1, \{r@b\})\})} & \equiv\\
\mathmob{b(h, C, \{ r: (\nullr, (\truek, \{\selfk : r\}, TimeServer))\},
     \launchT(\{\selfk : r\}, \epsilon, \nullr), \emptyset) \conc  \calA} &\\
     \mathmob{(\Ans + \{r@b : h\}, \Sns + \{Time : (\alpha_1, \{r@b\})\})} & \eqdef
\end{greduce}

From now on we will denote $\mathmob{(\Ans + \{r@b : h\}, \Sns +
  \{Time : (\alpha_1, \{r@b\})\})}$ as $(\Ans', \Sns') =
\mathmob{R'}$.  We proceed unfolding the \launchT macro.

\begin{greduce}
\mathmob{b(h, C, \{ r: (\nullr, (\truek, \{\selfk : r\}, TimeServer)), \underline{t' : (t', \nullr)}\},
     \underline{(t', (\{\selfk : r\}, \epsilon), \nullr)}, \emptyset) \conc  \calA, \underline{R'}} & \reduces
\end{greduce}

Since the code for \maink is empty, we may apply rule \rulename{End} to terminate the thread. 

\begin{greduce}
\mathmob{b(h, C, \{ r: (\nullr, (\truek, \{\selfk : r\}, TimeServer)), t' : (t',
  \nullr)\}, \underline{\notify(t')},  \emptyset) \conc  \calA, R'} &\reduces
\end{greduce}

To notify the threads suspended on $\mathmob{t'}$ we apply rule
\rulename{Notify}.  Note that there are no suspended threads, and thus
the rule does not wake any thread.

\begin{greduce}
\mathmob{b(h, C, \{ r : (\nullr, (\truek, \{\selfk : r\}, TimeServer)), t' : (t',
  \nullr)\}, \underline{\zerokt},  \emptyset) \conc  \calA, R'} &
\end{greduce}

We denote this network configuration as $\calA', \ns'$, and launch the
$\textsf{TimeClient}$ from listing \ref{lst:lang:timec} script onto it. 
Note that, in order
to avoid using to many identifiers, we may repeat some of the ones
used for references, since they are lexically bound to the agent that
hosts them.


Let $\mathmob{a'}$ be the key for the agent executing the client
script at host $\mathmob{h'}$, and $\mathmob{P'}$ be the code for the \maink\ method of the
$\mathmob{TimeClient}$ agent.  The initial state of the network with
the spawning of the new agent is:

\begin{greduce}
   \mathmob{\underline{a'(h', C', \emptyset, \launchT(\emptyset,  hosts = \newk\ Array(\nullk, 0);\ ...;\ \exitk,
     \nullr),\emptyset)}} \conc  \calA', \ns' & \eqdef
\end{greduce}

where $\mathmob{C' = C_0 + \{TimeClient : (\truek, hostList, \{\maink
  : (\epsilon, P')\}, C_0, \epsilon)\}}$ is the code repository for
agent $\mathmob a'$ (result of the \codeclo function) that holds in
$\mathmob C_0$ the code for the $\mathmob{Array}$ class.  Unfolding
the \launchT macro we have:

\begin{greduce}
   \mathmob{a'(h', C', \underline{\{t : (t, \nullr)\}}, \underline{(t, (\emptyset, hosts = \newk\ Array(\nullk, 0);\ ...;\ \exitk),
     \nullr)},\emptyset) \conc \calA', \ns'} \reduces
\end{greduce}

We now jump to line 23 and skip the creation of the  $\mathmob{hosts}$ array.
Assume that  at the time 
of the creation of the agent, variable $\mathmob{hosts}$ is bound to a
reference $\mathmob{r}$, that holds the entire array. 
We denote the heap of $\mathmob{a'}$ at this point as $\mathmob{H}$,
and the set of bindings of the thread as $\mathmob{B}$ that contains
$\mathmob{\{hosts : r\}}$. In line 23 we find the creation of the
agent:

\begin{greduce}
   \mathmob{a'(h', C', \underline{H}}, 
\mathmob{(t, (\underline{B}, \underline{x = \newk\ TimeClient(hosts)};\ \exitk), \nullr), \emptyset) \conc \calA', \ns'} & \reduces
\end{greduce}

We now apply rule \rulename{NewAgent} to create the agent.  For that
we introduce a new agent key, $\mathmob{b'}$. The values given to the
attributes of the class must be cloned. 
We thus perform the cloning of the reference bound to $\mathmob{hosts}$
and collect the code required by the methods of the
$\mathmob{Array}$ class (of which $\mathmob{hosts}$ is an instance).

\begin{greduce}
\mathmob{  \evalseq(H, t, B, hosts)} & = & \mathmob{r}\\
\mathmob{ \copyseq_{a'b'}(C', H, r)} & = & \mathmob{(C_0, H', r')}\\
\mathmob{  \codein(C', \maink() \{P'\})} & = &\emptyset
\end{greduce}

$\mathmob{H'}$ and $\mathmob{C_0}$ are respectively the heap and code
closure for $\mathmob{r'}$.
Thus, the heap of the new agent will be $\mathmob{H' + \{ r'' :
  (\nullr, (\truek, \{hostList : r'\}, TimeClient))\}}$, where $\mathmob{r''}$ is
the reference that holds the agent's closure.
Since $\mathmob{hosts}$ holds an array, the only code required by the
reference associated to it in the new agent is the $\mathmob{Array}$
class, kept in the $\mathmob{C_0}$ code repository.
Thus, the code repository for the new agent will be $\mathmob{C_0 +
  \{TimeClient : (\truek, hostList, \{\maink : (\epsilon, P')\}, C_0,
  \epsilon)\} = C'}$, which contains the code for $\mathmob{Array}$
and $\mathmob{TimeClient}$. Note that the methods of class
$\mathmob{TimeClient}$ do not require any extra code.  The resulting
state is:

\begin{greduce}
   \mathmob{a'(h', C', H, (t, (\underline{B + \{x : r''@b'\}}, \underline{\exitk}), \nullr), \emptyset)} \conc\\
   \mathmob{\underline{b'(h', C', H' + \{ r'' : (\nullr, (\truek, \{hostList : r'\}, TimeClient))\},
     \launchT(\{\selfk : r''\}, P', \nullr), \emptyset)}} \conc &\\
     \mathmob{\calA', (\underline{\Ans' + \{r''@b' : h'\}}, \Sns')} & \reduces \equiv
\end{greduce}

We then terminate the execution the client script
(agent $\mathmob{a'}$) and garbage collect it to obtain:

\begin{greduce}
   \mathmob{b'(h', C', H' + \{ r'' : (\nullr, (\truek, \{hostList : r'\}, TimeClient))\},
     \launchT(\{\selfk : r''\}, P', \nullr), \emptyset)} \conc &\\
     \mathmob{\calA', (\Ans' + \{r''@b' : h'\}, \Sns')} & \eqdef
\end{greduce}

Unfolding once again the \launchT macro we obtain the state of the
client prior to its interaction with the $\mathmob{Time}$ service
provider:

\begin{greduce}
   \mathmob{b'}(\mathmob{h', C', H' + \{ r'' : (\nullr, (\truek,
     \{hostList : r'\}, TimeClient)), \underline{t' : (t', \nullr)}\}},\\
    \mathmob{ \underline{(t', (\{\selfk : r''\}, P'), \nullr)}, \emptyset)} \conc &\\ \calA', \mathmob{(\Ans' + \{r''@b' : h'\}, \Sns')} & 
\end{greduce}

Remember that ${\small \mathmob{\calA' = b(h, C', \{ r': (\nullr, (\truek,
  \{\selfk : r'\}, \emptyset, TimeServer)), t' : (t',
  \nullr)\}, \underline{\zerokt},  \emptyset) \conc  \calA}}$.
To perform the interaction between both agents we need to expose both
their states from the network.  To present a less extensive state we
denote the heap of each agent as:

\begin{greduce}
\mathmob{H''} & = & \mathmob{\{ r': (\nullr, (\truek, \emptyset, TimeServer)),  t' : (t',   \nullr)\}} \\
\mathmob{H'''} & =&   \mathmob{H' + \{r'' : (\nullr, (\truek, \{hostList : r'\}, TimeClient)),
  t' : (t', \nullr) \}}\\
\end{greduce}

\vspace{-22pt}
Thus, by using the congruence rules we can obtain the following
configuration of the network with both agents exposed and associated:

\begin{greduce}
(\overbrace{\mathmob{b'(h', C', H''',
    (t', (\{\selfk : r''\}, \underbrace{\mathmob{timeServer =
    \bindk(Time);\ ...}}_{\mathmob{P'}}), \nullr),
    \emptyset)}}^{Client} \conc  
\overbrace{\mathmob{b(h, C, H'', \zerokt,  \emptyset)}}^{Server}) \conc \calA,  & \\
 \mathmob{(\Ans' + \{r''@b' : h'\}, \Sns')} &   \reduces
\end{greduce}

We are now going to execute  some of the code in
$\mathmob{P'}$, the code for the \maink\ method of the client. 
We begin in line 3, where we apply rule \rulename{BindAny}. We know
that $\mathmob{\{r@b : h'\}}$ is present in the resolver, thus
$\mathmob{r@b}$ is the result of the binding operation.

\begin{greduce}
(\mathmob{b'(h', C', H''',
     (t', (\{\selfk : r'', \underline{timeServer : r@b}\},
     \underline{i = hostList.iterator()};\ ...), \nullr), \emptyset)}
     \conc  & \\
\mathmob{b(h, C, H'', \zerokt,  \emptyset)}) \conc \calA,
\mathmob{(\Ans' + \{r''@b' : h'\}, \Sns')} &  \equiv \reduces^*
\end{greduce}

Next we proceed with the execution of agent $\mathmob{b'}$.
In
order to apply rules over this agent, we have to disassociate it from
agent $\mathmob{b}$. For that we use rule \rulename{Agent-Assoc}. 

\begin{greduce}
\underline{\mathmob{b'(h', C', H''',
     (t', (\{\selfk : r'', timeServer : r@b\},
     i = hostList.iterator();\ ...), \nullr), \emptyset)}}
     \conc  & \\
\mathmob{b(h, C, H'', \zerokt,  \emptyset)} \conc \calA,
\mathmob{(\Ans' + \{r''@b' : h'\}, \Sns')} &  \equiv \reduces^*
\end{greduce}

Now we have a state from which we may execute agent $\mathmob{b'}$.
In its code we skip the instructions until line 9, where we find a
\gok.  Consider that the current state is:

\begin{greduce}
\underline{\mathmob{b'(h', C', H''',
     (t', (B',  \gok(hostName)};\ ...), \nullr),
     \emptyset) \conc  
\mathmob{b(h, C, H'', \zerokt,  \emptyset)}} \conc \calA, \mathmob{(\Ans' + \{r''@b' : h'\}, \Sns')} &  \reduces
\end{greduce}

where we denote the updated bindings as $\mathmob{B'}$, assuming that
$\mathmob{B'(hostName) = h''}$.
We now apply rule \rulename{Go} to migrate the agent to $\mathmob{h''}$:

\begin{greduce}
\mathmob{b'(\underline{h''}, C', H''',
  (t', (B',   \underline{time = timeServer.getTime()};  ...),
     \nullr), \emptyset)} \conc  
\mathmob{b(h, C, H'', \zerokt,  \emptyset)} \conc \calA, & \\
\mathmob{(\underline{\Ans' + \{r''@b' : h''\}}, \Sns')} & \equiv
\end{greduce}

Next, we have a method invocation on the $\mathmob{timeServer}$ agent.
Since  $\mathmob{B'(timeServer) = r@b}$, and  the agent
running the thread is $\mathmob{b'}$,  the invocation
is remote.  We have thus to apply rule \rulename{RemoteInvoke}.
Neither the heap, nor the code repository of the target agent are
modified, since the method as no parameters.  Before applying it, we
remember the rule for remote invocation.

\ppcnnrule{
   \rulename{RemoteInvoke}
}{
  \evalseq(H, t, B, \tilde v) = \tilde u \quad   
  B(o)= \at{r'}{b} \quad
  H'(\at{r'}{b}) = (\_, (\truek, B', X)) 
}{
  \copyseq_{ab}(C, H, \tilde u) = (C', H'', \tilde u') \quad
   r''@a \in\ \HeapRef\  \fresh
}{
  (a(h, C, H, (t, (B,  x = o.m(\tilde v)\ ; P) :: Q, r) \conc\ T, W) \conc\ b(h', C, H', T', W')) \conc \calA, \ns
}{
  (a(h, C, H + \{r'' : (t, \nullr)\}, T, W + \{r'' : (t, (B + \{x :
  r''\}, P) :: Q, r) \}) \conc\ 
}{
  b(h', C + C', H' + H'', \launchT(\{\selfk: r', \tilde x : \tilde
  u'\}, x = \selfk.m(\tilde x) ; \returnk (x), r''@a) \conc\  T', W')) \conc \calA, \ns
}

Before applying the rule we have to associate the agents that will
take part on the communication. Thus, applying rule
\rulename{AgentAssoc} we have:

\begin{greduce}
(\mathmob{b'(\underline{h''}, C', H''',
  (t', (B',   \underline{time = timeServer.getTime()};  ...),
     \nullr), \emptyset)} \conc  
\mathmob{b(h, C, H'', \zerokt,  \emptyset)}) \conc \calA, & \\
\mathmob{(\underline{\Ans' + \{r''@b' : h''\}}, \Sns')} & \reduces
\end{greduce}

We may now apply rule \rulename{RemoteInvoke} to 
create the new thread in the target agent, and suspend the calling
thread, waiting for the result.

\begin{greduce}
(\mathmob{b'(h'', C', \underline{H''' + \{r''' : (t', \nullr)\}}, \underline{\zerokt}, \underline{\{r''' :
     \{(t', (B' + \{ time : r'''\},  command = ...;), \nullr)\}}\})}
 \conc &\\
\mathmob{b(h, C, H'', \underline{\launchT(\{\selfk : r\}, x =
    \selfk.getTime() ; \returnk (x), r'''@b')} \conc \zerokt,
  \emptyset)}) \conc  \calA, &\\
\mathmob{(\Ans' + \{r''@b' : h''\}, \Sns')} & \equiv \eqdef
\end{greduce}

Now we proceed the execution in agent $\mathmob{b}$ to execute the
method. To have a state on which we can apply rules over
$\mathmob{b}$, we have to disassociate the agents and commute their
position. 

\begin{greduce}
\underline{\mathmob{b(h, C, H'', \launchT(\{\selfk : r\}, x =
    \selfk.getTime() ; \returnk (x), r'''@b') \conc \zerokt,  \emptyset)}} \conc &\\
\underline{\mathmob{b'(h'', C', H''' + \{r''' : (t', \nullr)\}, \zerokt, \{r''' :
     \{(t', (B' + \{ time : r'''\},  command = ...;), \nullr)\}\})}}
 \conc    \calA, &\\
\mathmob{(\Ans' + \{r''@b' : h''\}, \Sns')} & \equiv \eqdef
\end{greduce}

From now on we denote $\mathmob{(\Ans' + \{r''@b' : h''\}, \Sns')}$ as
$\mathmob{\ns''}$. Unfolding $\launchT$ we create a new thread
associated to a new thread reference ($\mathmob{t''})$.

\begin{greduce}
\mathmob{b(h, C, \underline{H'' + \{t'' : (t'', \nullr)\}} , \underline{(t'', (\{\selfk : r\}, x = \selfk.getTime() ; \returnk (x)), r'''@b')} \conc \zerokt,  \emptyset)} \conc &\\
\mathmob{b'(h'', C', H''' + \{r''' : (t', \nullr)\}, \zerokt, \{r''' :
     \{(t', (B' + \{ time : r'''\},  command = ...;), \nullr)\}\})} \conc \calA, \underline{\ns''} & \reduces
\end{greduce}

We now apply rule \rulename{LocalInvoke} to perform the local invocation in $\mathmob{b}$.
We know that:

\begin{greduce}
\mathmob{H''(r)} & = & \mathmob{(\nullr, (\truek, \emptyset, TimeServer))} \\
\mathmob{C(TimeServer)} & = & \mathmob{(\truek, \epsilon, \{\maink : (\epsilon,
     \epsilon),  getTime : (\epsilon, P)\}, \emptyset, Time)}
\end{greduce}

Thus, $\mathmob{P}$ is the code to execute.
We create a new reference $\mathmob{r'}$ to hold the result of the
method, suspend the current thread on it, and create a new thread,
associated to the same reference as the current ($\mathmob{t''}$), to execute the method.

\begin{greduce}
\mathmob{b(h, C, H'' + \{t'' : (t'', \nullr), \underline{r' : (t'', \nullr)}\},
  (t'', \underline{(\emptyset, P)}, r'), \underline{\{r' : \{(t'', (\{x : r'\}, \returnk (x)), r'''@b')\}\}})} \conc &\\
\mathmob{b'(h'', C', H''' + \{r''' : (t', \nullr)\}, \zerokt, \{r''' :
     \{(t', (B' + \{ time : r'''\},  command = ...;), \nullr)\}\})} \conc   \calA, \ns'' & \reduces
\end{greduce}

The body of the method ($\mathmob{P}$) executes until it reaches a
\returnk instruction, that returns the value for the time ($\mathmob{c}$). Consider
that $\mathmob{B''}$ are the updated bindings of the thread executing
$\mathmob{getTime}$, and that $\mathmob{B''(x) = c}$.

\begin{greduce}
  \mathmob{b}(\mathmob{h, C, H'' + \{t'' : (t'', \nullr), r' : (t'', \nullr)\}}, & \\
  \mathmob{(t'', (\underline{B''}, \underline{\returnk (x)}), r'), \{r' : \{(t'', (\{x : r'\}, \returnk (x)), r'''@b')\}\},  \emptyset}) \conc &\\
\mathmob{b'}(\mathmob{h'', C', H''' + \{r''' : (t', \nullr)\}}, 
\mathmob{\zerokt, \{r''' :
     \{(t', (B + \{ time : r'''\},  command = ...;), \nullr)\}\}}) \conc \calA, \ns'' & \reduces
\end{greduce}

We will now apply rule \rulename{Return}. The value is placed on the
correspondent reference.

\begin{greduce}
\mathmob{b(h, C, H'' + \{t'' : (t'', \nullr), \underline{r' : (\nullr, c)}\},
  \notify(r'), \{r' : \{(t'', (\{x : r'\}, \returnk (x)), r'''@b')\}\})} \conc &\\
\mathmob{b'(h'', C', H''' + \{r''' : (t', \nullr)\}, \zerokt, \{r''' :
     \{(t', (B + \{ time : r'''\},  command = ...;), \nullr)\}\})}
   \conc   \calA, \ns'' & \reduces
\end{greduce}

Applying rule \rulename{Notify}.  Note that, from the definition of \run in section \ref{sec:mobam}, we have that:

\begin{step}
\mathmob{\run(\{(t'', (\{x : r'\}, \returnk (x)), r'''@b')\}) = (t'', (\{x : r'\}, \returnk (x)), r'''@b')}
\end{step}

The resulting state is thus:

\begin{greduce}
\mathmob{b(h, C, H'' + \{t'' : (t'', \nullr), r' : (\nullr, c)\},
 \underline{(t'', (\{x : r'\}, \returnk (x)), r'''@b')}, \underline{\emptyset})} \conc &\\
\mathmob{b'(h'', C', H''' + \{r''' : (t', \nullr)\}, \zerokt, \{r''' :
     \{(t', (B + \{ time : r'''\},  command = ...;), \nullr)\}\})}
     \conc&\\  \calA, \ns'' & \equiv \reduces
\end{greduce}

To return the result back to the calling agent we have to apply rule
\rulename{RemoteReturn}. Before we need to associate the agents, we
skip over this part.  The value to be placed in the heap of the calling
agent $\mathmob{b'}$, the target of \returnk, must be a clone of the
value in the heap of $\mathmob{b}$. The operation is performed by the
\copyone function.

\begin{greduce}
  \mathmob{\evalone(H'' + \{t'' : (t'', \nullr), r' : (\nullr, c)\}, t'', \{x : r'\}, x)} & = & \mathmob{c} \\
  \mathmob{\copyone_{bb'}(C, H + \{t'' : (t'', \nullr), r' : (\nullr, c)\}, c)} &=&  \mathmob{(\emptyset, \emptyset, c)}
\end{greduce}

Thus, we obtain:

\begin{greduce}
(\mathmob{b(h, C, H'' + \{t'' : (t'', \nullr), r' : (\nullr, c)\}, \underline{\zerokt}, \emptyset)} \conc &\\
\mathmob{b'(h'', C', \underline{H''' + \{r''' : (\nullr, c)\}}, \underline{\notify(r''')},
  \{r''' : \{(t', (B + \{ time : r'''\},  command = ...;),
  \nullr)\}\})}) \conc&\\  \calA, \ns'' & \eqdef \reduces &
\end{greduce}

We may now apply rule \rulename{Notify} to wake the calling thread.
Remember that in order to apply the rule on agent $\mathmob{b'}$ we
need to disassociate and commute the agents.

\begin{greduce}
\mathmob{b'(h'', C', H''' + \{r''' : (\nullr, c)\}, 
\underline{(t', (B + \{ time : r'''\},  command = ...;), \nullr)}, \underline{\emptyset})}
 \conc &\\
\mathmob{b(h, C, H'' + \{t'' : (t'', \nullr), r' : (\nullr, c)\},
  \underline{\zerokt}, \emptyset)} \conc 
  \calA, \ns'' & \reduces
\end{greduce}

Thus, the thread that performed the method invocation resumes its
execution, with a binding for the value ($\mathmob{c}$), through
reference $\mathmob{r'''}$.  We conclude here the partial execution of
our example. The server agent enters a state of idleness waiting for new
requests, while the agent will execute the \texttt{setTimeApplication}
binary program to set the local clock.

\section{Conclusions}
\label{sec:conclusions}

In this report we have presented the syntax and semantics for the core
of a language for programming mobile agents, named \mob. A new report,
focusing on the encoding of the semantics of this language into a
process calculus is currently being prepared.

The \mob core-language compiler and run-time system are implemented.
The first is an implementation of the encoding to be presented in the
next report, and the second an extension to the distributed \ptyco
run-time to allow the execution of \mob computations.

The run-time is currently being extended with primitives for
interaction with external services. This will allow \mob to act as a
coordination language for mobile agents that interact with web
services for: recognition/execution of programs in several high-level
languages, building itineraries through external search engines,
database transactions, and network communication through known
protocols, such as SMTP, FTP, or HTTP.

Future plans also include an integrated tool for programming,
debugging and monitoring agents.

\bibliographystyle{plain}
\bibliography{refs}

\end{document}
